\documentclass[conference,anonymous]{IEEEtran}
\IEEEoverridecommandlockouts
\usepackage{cite}
\usepackage{amsmath,amssymb,amsfonts}
\usepackage{algorithmic}
\usepackage{graphicx}
\usepackage{textcomp}
\usepackage{xcolor}
\usepackage{enumitem}
\usepackage{todonotes}
\usepackage{hyperref}
\usepackage{comment}

\def\BibTeX{{\rm B\kern-.05em{\sc i\kern-.025em b}\kern-.08em
    T\kern-.1667em\lower.7ex\hbox{E}\kern-.125emX}}
\begin{document}

\title{Strategic Bidding Wars in On-chain Auctions}

\author{%
    \IEEEauthorblockN{Fei Wu\IEEEauthorrefmark{1}, Thomas Thiery\IEEEauthorrefmark{2}, Stefanos Leonardos\IEEEauthorrefmark{1} and Carmine Ventre\IEEEauthorrefmark{1}}
    \IEEEauthorblockA{\IEEEauthorrefmark{1}Department of Informatics, King's College London, United Kingdom}
    \IEEEauthorblockA{\IEEEauthorrefmark{2}Robust Incentives Group, Ethereum Foundation}
    \IEEEauthorblockA{\IEEEauthorrefmark{1}\{fei.wu, stefanos.leonardos, carmine.ventre\}@kcl.ac.uk, \IEEEauthorrefmark{2}thomas.thiery@ethereum.org. }    
}
\IEEEoverridecommandlockouts

\maketitle

\begin{abstract} 
The Ethereum block-building process has changed significantly since the emergence of Proposer-Builder Separation. Validators access blocks through a marketplace, where block builders bid for the right to construct the block and earn MEV (Maximal Extractable Value) rewards in an on-chain competition, known as the MEV-boost auction. While more than 90\% of blocks are currently built via MEV-Boost, trade-offs between builders' strategic behaviors and auction design remain poorly understood.\par
In this paper we address this gap. We introduce a game-theoretic model for MEV-Boost auctions and use simulations to study different builders' bidding strategies observed in practice. We study various strategic interactions and auction setups and evaluate how the interplay between critical elements such as access to MEV opportunities and improved connectivity to relays impact bidding performance. Our results demonstrate the importance of latency on the effectiveness of builders' strategies and the overall auction outcome from the proposer's perspective.

\end{abstract}


\section{Introduction}

Ethereum's transition to Proof-of-Stake (PoS) introduced \emph{Proposer-Builder Separation} (PBS), aimed at reducing computational demands and enhancing decentralization for Ethereum validators. PBS segregates the block production process and assigns the complex task of transaction ordering and block construction for \emph{Maximal Extractable Value} (MEV) extraction to block builders, while validators focus on selecting and publishing the optimal builder-created block. To protect builders from MEV theft and ensure fair practices until enshrined PBS \cite{epbs} is implemented into the consensus layer of the Ethereum protocol, Flashbots and Ethereum Foundation developed \emph{MEV-Boost} \cite{FlashbotsmevBoost}, as an out-of-protocol solution. \par

On the Ethereum 2.0 beacon chain \cite{beacon}, time progresses in 12-second slots \cite{gasper,ghost,casper}. In each slot, a validator is randomly selected as the proposer to propose a block, while other validators are responsible for providing attestations. According to the honest validator specification \cite{honestvali}, the proposer is expected to propose a block promptly at the beginning of the 12-second slot. MEV-Boost enables the proposer to outsource block construction to builders on a marketplace via trusted intermediaries known as relays. This marketplace hosts a competition, termed the ``MEV-Boost auction," in which builders vie for the right to construct the block. The winner of this auction, which is randomly terminated by the proposer, typically spanning the entire 12-second interval, will have their block selected and proposed.\par
Within every auction cycle, builders submit valid blocks alongside their bids to relays. Bid values represent block rewards, including priority fees from user transactions in the public mempool and payments from searchers for bundle inclusion. Searchers detect MEV opportunities (e.g., arbitrages, sandwich attacks, liquidations, and cross-domain MEV \cite{qin2022quantifying, obadia2021unity}) and strategically bundle their transactions with existing user transactions to capitalize on them. Searchers' transaction bundles, termed \emph{exclusive orderflow} (EOF), are sent directly and privately either to multiple builders or to a single integrated searcher-builder. Builders, then, pack these bundles and other user transactions into maximally profitable blocks, forwarding these blocks with their bids to relays.\par
Relays act as trusted facilitators between builders and the proposer, by validating received blocks, selecting the winning block with the highest bid, and forwarding only the winning block header to the proposer when the proposer calls {\tt getHeader} to request the block \cite{relay}. This ensures that proposers cannot steal block content but can commit to proposing the block by signing the respective block header. Unfortunately, the default relay implementation adds significant \emph{latency} to MEV-boost auctions. 
This incentivizes vertical integration which in turn reduces proposer fees and compromises network security \cite{Chi23}. To address these shortcomings, \emph{optimistic relays} assume that received blocks are valid in the short-term, and defer actual validation to a later time \cite{Neu23}. \par
MEV-Boost acts as a relay aggregator and allows both builders and proposers to connect with multiple relays, enabling them to increase their chances of winning bids and optimize their profits, respectively \cite{FlashbotsmevBoost}. The auction concludes as soon as the proposer selects and signs the header. Upon the proposer returning the signed header to the relay, the relay publishes the full block to the p2p network. Presently, over 90\% of Ethereum blocks are constructed via MEV-Boost \cite{mevBoostpics}.\par

Despite the dominance of MEV-Boost auctions in PoS-Ethereum and the keen interest of the academic and industry communities to understand this complex block-building process, there is notably scarce literature and limited practical insight on both bidding tactics and auction design.
In this paper, we fill this gap by exploring the intricate interplay between strategic bidding behaviors and MEV-Boost auction design features. We develop a game-theoretic model along multiple empirically observed strategies
(Section~\ref{sec:model}) and simulate MEV-Boost auctions under different strategic profiles and system configurations (Section~\ref{sec:sim}). We study the effectiveness of strategies under different scenarios (Section~\ref{sec:result} \&~\ref{sec:discuss}) and analyze the impact of important design decisions on auction outcome from the proposer's perspective (Section~\ref{sec:outcome}). We moreover compare the impact of EOF access and latency on builders' bidding performance (Section ~\ref{sec:eof}).
This paper makes the following contributions:
\begin{enumerate}[leftmargin=*,wide=0pt]
    \item 
    Inspired by \cite{model}, we propose a 
    game-theoretic model for MEV-Boost auctions, where block builders compete as bidders for the block construction right 
    and block proposers act as auctioneers. 
    The model encompasses various auction settings affecting 
    global dynamics and 
    individual participants.
    
    \item We conduct a comprehensive analysis of builders' strategic behaviors in MEV-Boost auctions. We introduce four 
    bidding strategies based on recent empirical observations \cite{bbp, cancellation} -- naive, adaptive, last-minute, and bluff -- and use agent-based modeling (ABM) to simulate scenarios in which builders using these strategies compete according to our model.    
    \item 
    We demonstrate how factors such as latency, exclusive orderflow, builders' strategic decisions, and auction design aspects, including auction termination times and relay types, impact builders' bidding performance and strategy effectiveness across various scenarios. We assess the efficiency of the auction from the proposer's perspective and provide insights into the impact of latency on the expected outcomes. Amongst others, our results demonstrate the importance of reduced latency in expected performance, and explain why both builders and proposers favor optimistic relays.

\end{enumerate}

\section{MEV-Boost Auction Model}
\label{sec:model}

We consider a set of $N = \{1, \ldots, n\}$ builders competing in the MEV-Boost auction game. Each builder, indexed by $i$, employs a bidding strategy $s$ which can be described as a function $\beta_s: X \to \mathbb{R}_+$ so that the bid of player $i$ at time $t$ is $b_{i,t} =\beta_s(x_{i,t})$, where $x_{i,t} \in X$ represents a 
vector of input variables 
at time $t \geq 0$. These inputs are discussed next, 
where the terms \emph{player} and \emph{builder} are used interchangeably. 

\begin{itemize}[leftmargin=*,wide=0pt]
\item \textbf{Public signal} $P(t)${\bf.} The public signal represents priority fees and payments from transactions in the public mempool, accessible to all builders. Users submit new transactions with priority fees as the auction advances. We model this as a compound Poisson process, where the number of transactions, denoted by $N(t)$, in the public mempool up to time $t$, follows a Poisson distribution with rate $\lambda_p$, i.e., $N(t) \sim \text{Poisson}(\lambda_p \cdot t)$. The MEV value, $V_j$, of each transaction, $j$, is a random variable drawn from a log-normal distribution, i.e., $V_j \sim \text{Log-normal}(\xi_1, \omega_1)$. The public signal, denoted as $P(t)$ and identical for all builders, is the cumulative sum of MEV values of $N(t)$ transactions, given by the equation:
\begin{equation}\label{eq:P}
P(t) = \sum\nolimits_{j=1}^ {N(t)} V_j.
\end{equation}


\item \textbf{Private signal} $E_i(t)${\bf.} The private signal denotes private transactions and payments secured from searchers, specifically, MEV of EOF. Typically, searchers send the same transaction bundle to multiple builders, indicating a common scenario where various builders frequently receive similar orderflow from the same group of searchers. Similar to the public signal, we model this as a compound Poisson process. We use a Poisson distribution with a rate $\lambda_e$ to determine the number of bundles, and a log-normal distribution to establish the value of each bundle. However, to account for the exclusiveness and correlation of orderflow among different players, we introduce EOF access probability, $\pi_i$, for each player $i \in N$, representing the player's probability of accessing each searcher bundle. These remain constant throughout the auction interval.
Using a similar notation, we write $N_i(t)$ for the number of bundles received by player $i$ up to time $t$, where $N_i(t)$ follows a Poisson process with rate $\lambda_e$, adjusted by the player's probability $\pi_i$, i.e., $N_i(t) \sim \text{Poisson}(\lambda_e \cdot t \cdot \pi_i)$. The MEV value, $O_j$, of bundle $j$ is randomly drawn from a log-normal distribution, i.e., $O_j \sim \text{Log-normal}(\xi_2, \omega_2)$. The private signal, $E_i(t)$, of player $i$ is
\begin{equation}\label{eq:Eit}
E_i(t) = \sum\nolimits_{j=1}^{N_i(t)} O_j.
\end{equation}
We also write $E(t)$ to denote the total value of all bundles at time $t$. Thus, the \emph{aggregated signal}, $S_i(t)$, of player $i$ at time $t$ and the \emph{total signal}, $S(t)$, in the auction at time $t$ can be calculated by combining \eqref{eq:P} and \eqref{eq:Eit}
\begin{align*}
S_i(t) = P(t) + E_i(t) \text{ and } 
S(t) = P(t) + E(t).        
\end{align*}
\end{itemize}
Since correlations between bid arrival times and bid values are positive \cite{schwarz2023time}, we assume these MEV opportunities are persistent and always profitable throughout the auction. We also assume the MEV of public transactions is universal for all builders and the payments secured from searchers are uniform between those builders who share the bundles.


\begin{itemize}[leftmargin=*,wide=0pt]
\item \textbf{Latency} $\Delta + \Delta_i${\bf.} The \emph{latency} causes a delay in the relay's acceptance of bids relative to the players' bid submissions. We introduce two components that constitute this latency: \emph{global delay}, influenced by the relay's processing speed, and \emph{individual delay}, determined by the player's network connectivity. Global delay, denoted as $\Delta$, quantifies the delays experienced by the relay in accepting these bids relative to their arrival. Individual delay, denoted as $\Delta_i$, represents the delays in the arrival of player $i$'s bids to the relay relative to their submission.

\end{itemize}
We assume both global and individual delays to be constant and known during the auction, and to only affect players' bidding actions. 

\begin{itemize}[leftmargin=*,wide=0pt]
\item \textbf{Profit margin} $pm_i${\bf.} 
The profit margin, $pm_i$, reflects player $i$'s $\in N$ unique risk tolerance and profit expectations.
\item \textbf{Current highest bid} $\max_{j \in N}\{b_{j,k}\}_{k \leq t}${\bf.} This variable represents the 
highest bid among all bids 
submitted by all builders up to time $t$. This information is known to all players. 
\end{itemize}

Beyond the strategy space and the inputs defined above, our model also captures the following elements.\par \smallskip
\noindent\textbf{Bid cancellations and honest proposer}: Bid cancellations are achieved by replacing existing bids with lower ones. Honest proposers typically make a single {\tt getHeader} request at the start of the slot and commit to the block with the highest bid at that moment. However, proposers can call {\tt getHeader} multiple times, allowing them to initially select a block with a higher bid and maintain this choice even if a subsequent {\tt getHeader} request reveals a lower bid, rendering the bid cancellation ineffective. In our model, we assume proposers are honest, or equivalently that bid cancellation is effective. Builders are aware of the proposer's honesty and expect the proposer to terminate the auction at the beginning of the slot.

\begin{table}[t]
\centering
\caption{Bidding Strategies}
\label{tab:bidding_strategies}
\begin{tabular}{|p{1.5cm}|p{6.5cm}|}
\hline
\textbf{Strategy} & \textbf{
Bid value $b_{i,t}$ at time $t\leq T$} \\ \hline
Naive & $\beta_{\text{naive}}(x_{i,t})= v_i(t)_+$ \\ \hline
Adaptive & $\beta_{\text{ada}}(x_{i,t})= \min{\{v_i(t),\max_{j\in N} {\{b_{j,k}: k\leq t\}}+\delta\}_+}$ \\ \hline
Last-minute & $\beta_{\text{last}}(x_{i,t})= v_i(t)_+\times \mathbf{1}\{t\ge \theta\}$ \\ \hline
Bluff & $\beta_{\text{bluff}}(x_{i,t}) = b_{i,\text{bluff}} \times \mathbf{1}\{t < \theta\} + v_{i}(t)_+ \times \mathbf{1}\{t \geq \theta\}$ \\ \hline
\end{tabular}
\vspace{-0.3cm}
\end{table}

\smallskip \noindent {\bf Time progression.}
The MEV-Boost auction proceeds in continuous time over the interval [0, $T$], where $T$ denotes the time at which the winning bid is selected by the proposer. 
According to the protocol specifications, $T$ should be exactly equal to $12$ seconds, but, due to factors such as latency or proposer's strategic behaviors \cite{schwarz2023time}, the winning bid is typically selected around $T = 12$ seconds. We model $T$ using a Gaussian distribution with mean $12$ and standard deviation $\sigma$.


\smallskip \noindent {\bf Profits.} 
A builder submitting a valid block that wins the auction receives a profit equal to the difference between the total extractable MEV of the block and their bid. A losing builder has a profit of 0.
Thus, the profit, $u_i$, of player $i$ is
\vspace{-2mm}
\[u_i =
\begin{cases}
S_i(t_w) - b_{i,t_w} & \text{if $b_{i,t_w}=\max_{j \in N}\{b_{j,k}: k\leq T\}$} \\
0 & \text{otherwise}.
\end{cases}\]
where, $t_w$ denotes the submission time of the winning bid.

\smallskip \noindent {\bf Strategies.} Building on empirical 
studies of builders' bidding behaviors \cite{bbp, cancellation}, we formalize four bidding strategies: naive, adaptive, last-minute, and bluff. We will qualify players by their strategy; so, e.g.,  ``naive players'' are those playing the naive strategy. TABLE \ref{tab:bidding_strategies} summarizes the behavior of these players. 
%
The \emph{valuation}, $v_i(t)=S_i(t)-pm_i$, of player $i$ represents the highest bid player $i$ is able to place at time $t$ while ensuring a positive profit.

\begin{itemize}[leftmargin=*,wide=0pt]
\item \textbf{Naive} players follow a straightforward principle based on their valuation, $v_i(t)$. 
They aggressively update their bids as long as their aggregated signal surpasses their profit margin.

\item \textbf{Adaptive} players modify their bids in response to the observed behavior of other players. 
They monitor the current highest bid and place their bid if their valuation allows them to outbid by a small constant $\delta$ the current highest bid. If unable to outbid, adaptive players default to the naive strategy. 

\item \textbf{Last-minute} players 
reveal their valuation, as naive players, but only at the final possible moment before the auction is terminated. 
This strategy aims to minimize the window for other players, especially adaptive, to react to their bid, but carries the inherent risk of missing the submission window as the auction may end before their revelation. Empirical data suggests that this approach is being increasingly adopted by builders lately \cite{mevBoostpics}.

\item \textbf{Bluff} players intentionally place bids significantly higher than their valuation, denoted as $b_{i,\text{bluff}}$, before leveraging the bid cancellation feature to revert back to bidding their actual valuation toward the end of the auction. This strategy aims to compel other players, especially adaptive ones, to disclose their valuation. However, this strategy is risky as players may fail to cancel the bluff bids before the auction ends, potentially resulting in large negative profits if they win with the bluff bid.

\end{itemize}

\noindent\textbf{Revealing time.}
For last-minute and bluff players, we denote with $\epsilon$ the \emph{revealing time}, representing the gap between the revelation of the player's valuation and the expected auction termination ($T = 12$), i.e., 
the bid is accepted by the relay $\epsilon$ seconds before $T = 12$. The actual action of submitting that bid, however, is timed even earlier to compensate for potential delays in the bidding process. Thus, for player $i$, we define $\theta = 12 - \epsilon - \Delta - \Delta_i$, representing the time the player starts to bid their 
valuation.

\section{Experimental Setup}
\label{sec:sim}
\begin{figure}[t] \label{auction}
\centerline{\includegraphics[width=0.95\linewidth]{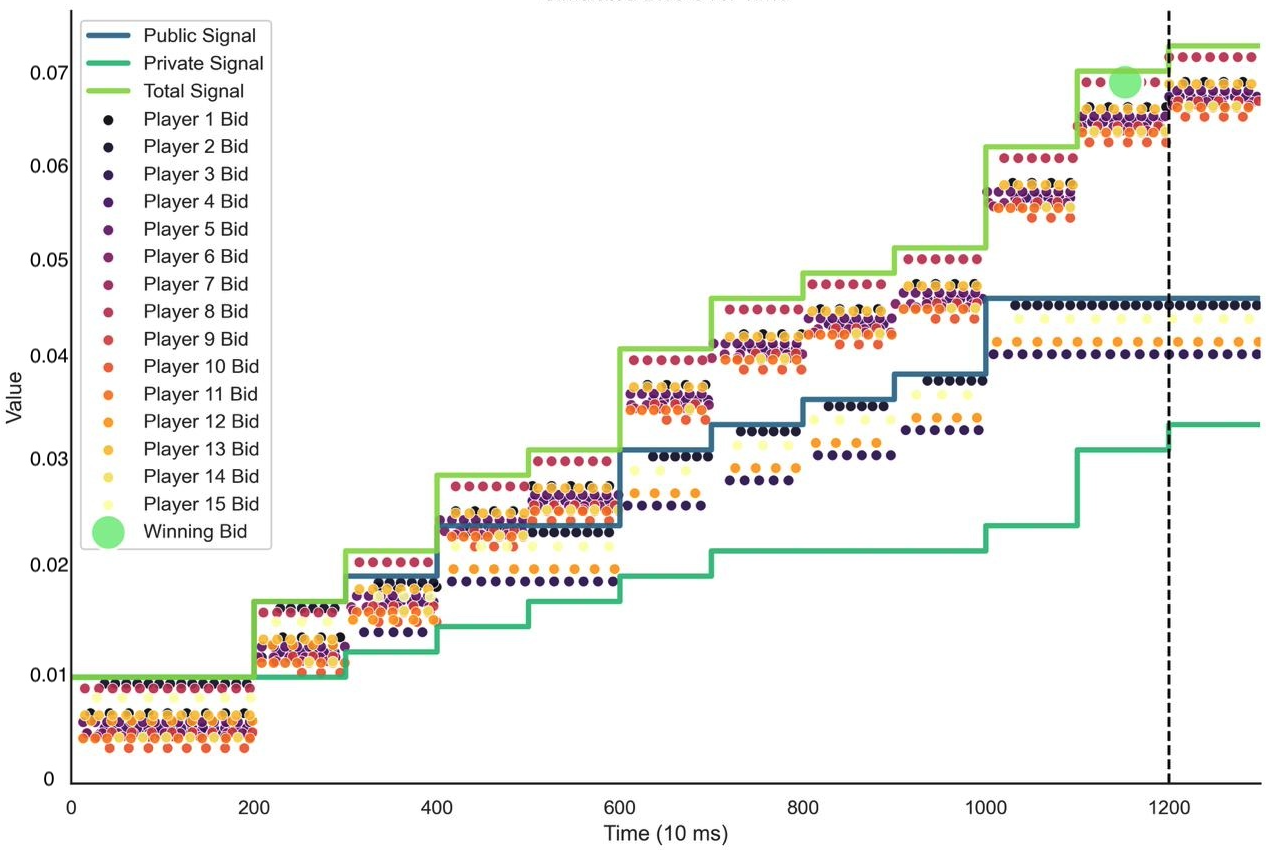}}
\caption{An example execution of our model to simulate one MEV-Boost auction. Players continue to bid based on their strategies and utilities until the auctioneer selects the winning bid and terminates the auction.}
\label{fig:auction}
\vspace{-0.2cm}\end{figure}

We use our model to simulate the bidding behaviors of players in MEV-Boost auctions via ABM (see Fig.~\ref{fig:auction}). We next discuss our experimental setup. 

\smallskip \noindent{\bf Simulation.}
For technical reasons, we assume the following. 
\begin{enumerate}[leftmargin=*,wide=0pt]
    \item \textbf{Discrete time}: In our simulation, time is represented through discrete time steps. 
    To approximate 
    the continuous flow of time 
    experienced in reality, one time step is 
    equivalent to 10 milliseconds (0.01 seconds).
    \item \textbf{Simultaneity}: Players observe and react to the current state of the auction at the same time. I.e., they take their bidding actions simultaneously at each time step of the simulation. In particular, this implies that adaptive players will react to the highest bid known at the previous time step. 
    \item \textbf{Signals:} The updates of both public 
    and private signals are accessible to the players at the same time, 
    which means that all the builders see new transactions in the public mempool and receive shared bundles from searchers at the same time.
\end{enumerate}


\smallskip \noindent{\bf Strategy profiles.}
According to the Nov 2023 data about MEV-Boost available on \cite{mevBoostpics},  
12 
builders share over 98\% of the total MEV-Boost block market. 
We thus choose 
two strategy profiles with 
12 players for our simulations. To fairly compare the effectiveness of strategies, the players are divided equally 
as follows. In 
%
\textbf{Profile 1}, we have 4 naive players, 4 adaptive players, and 4 last-minute players. In 
\textbf{Profile 2}, there are 4 naive players, 4 adaptive players, and 4 bluff players.

The common element in both profiles is the inclusion of naive players, who are assumed to be a constant presence in MEV-Boost auctions based on the existing examination of 
bid timing and values \cite{bbp}. By maintaining a consistent presence of naive and adaptive players in both profiles, we aim to establish a baseline for comparison between the relative effectiveness of the last-minute and bluff strategies.

We focus on three key performance metrics of the players: win rate, profit per win, and average profit (per auction simulation). By examining these metrics, we aim to gauge the overall performance of players within each strategy group, thereby determining the overall effectiveness of each strategy. We adjust the corresponding variables accordingly to investigate their impact on player performance and strategy effectiveness in both profiles. We conduct 10,000 auction simulations for each profile under each unique setting of variable values, ensuring the robustness of our simulation results. To prevent any observed differences in strategy effectiveness from being skewed by unequal distributions of other unrelated factors, for each auction simulation, we draw the EOF access probability $\pi_i$ and the profit margin $pm_i$ randomly from the same distributions for all players in both profiles. The model is calibrated against data derived from previous empirical analysis \cite{bbp}. For reproducibility, we provide the source code adopted for the simulations on \href{https://github.com/M1kuW1ll/MMASim}{github}. \footnote{We conducted experiments on 5 strategies but here present and study only 4 strategies, since the fifth 
 does not add new complexities.}.

\section{Strategy effectiveness}
\label{sec:result}

This section assesses strategy effectiveness across diverse influencing factors within the defined scenarios (profiles).

\smallskip \noindent \underline{\it Impact of latency.}
Both global delay and individual delay influence players' bidding actions in our model. 
However, as they capture different real-world aspects of auction design and builders' behavior in MEV-Boost auctions, we analyze their impact on strategy effectiveness separately.
To allow for a fair performance evaluation between strategies and assessment of individual delay impacts, we uniformly distribute individual delays across players within each strategy group in both profiles. Each strategy group has players with individual delays of 1-step (10ms), 2-step (20ms), 3-step (30ms), and 4-step (40ms). We fix the auction interval at $T = 12$ and the revealing time of last-minute and bluff players at the last auction step ($\epsilon = 0$) to isolate their effect on strategy performance.



\begin{figure}[t]
\centering
{\includegraphics[width=0.49\linewidth]{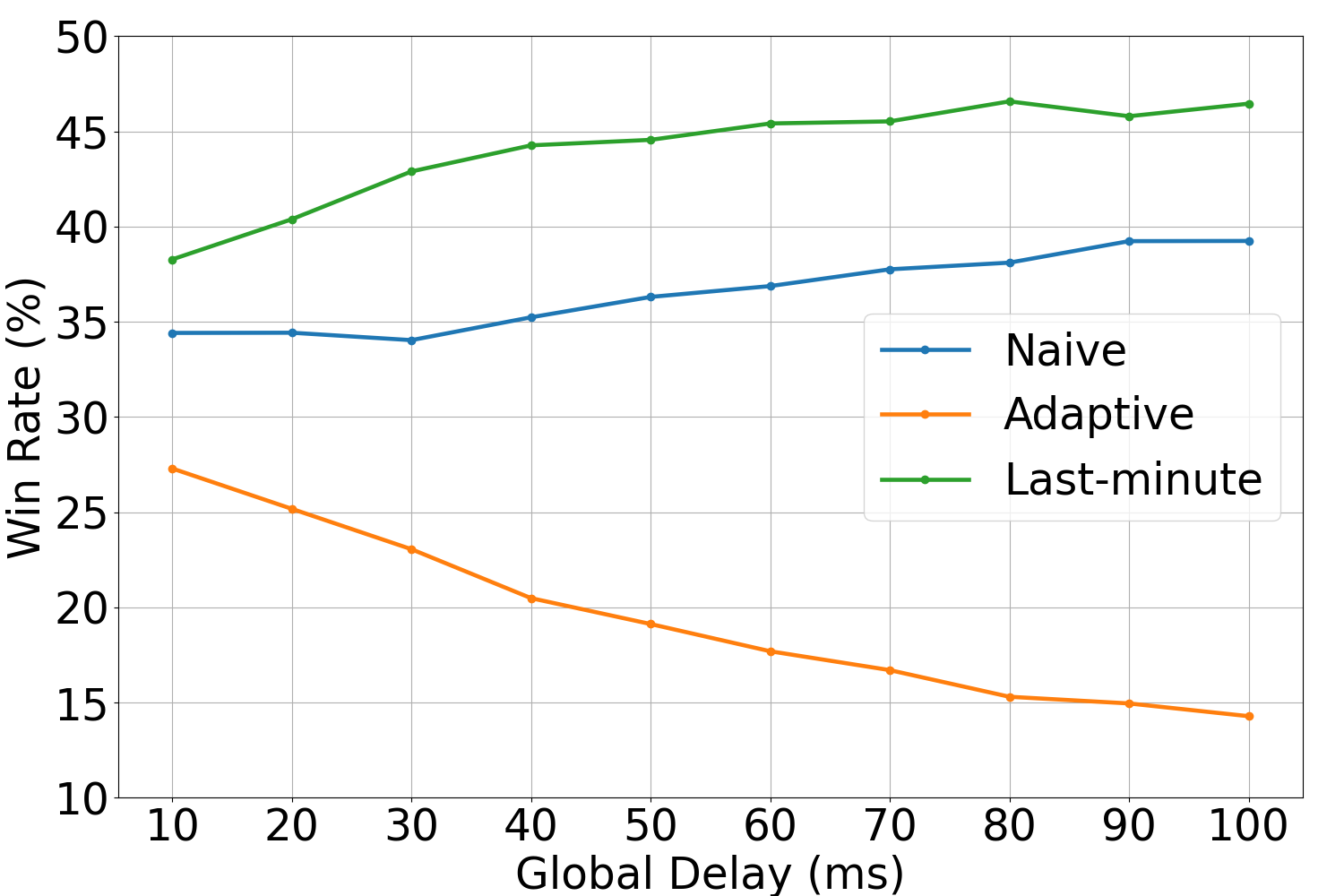}}
{\includegraphics[width=0.49\linewidth]{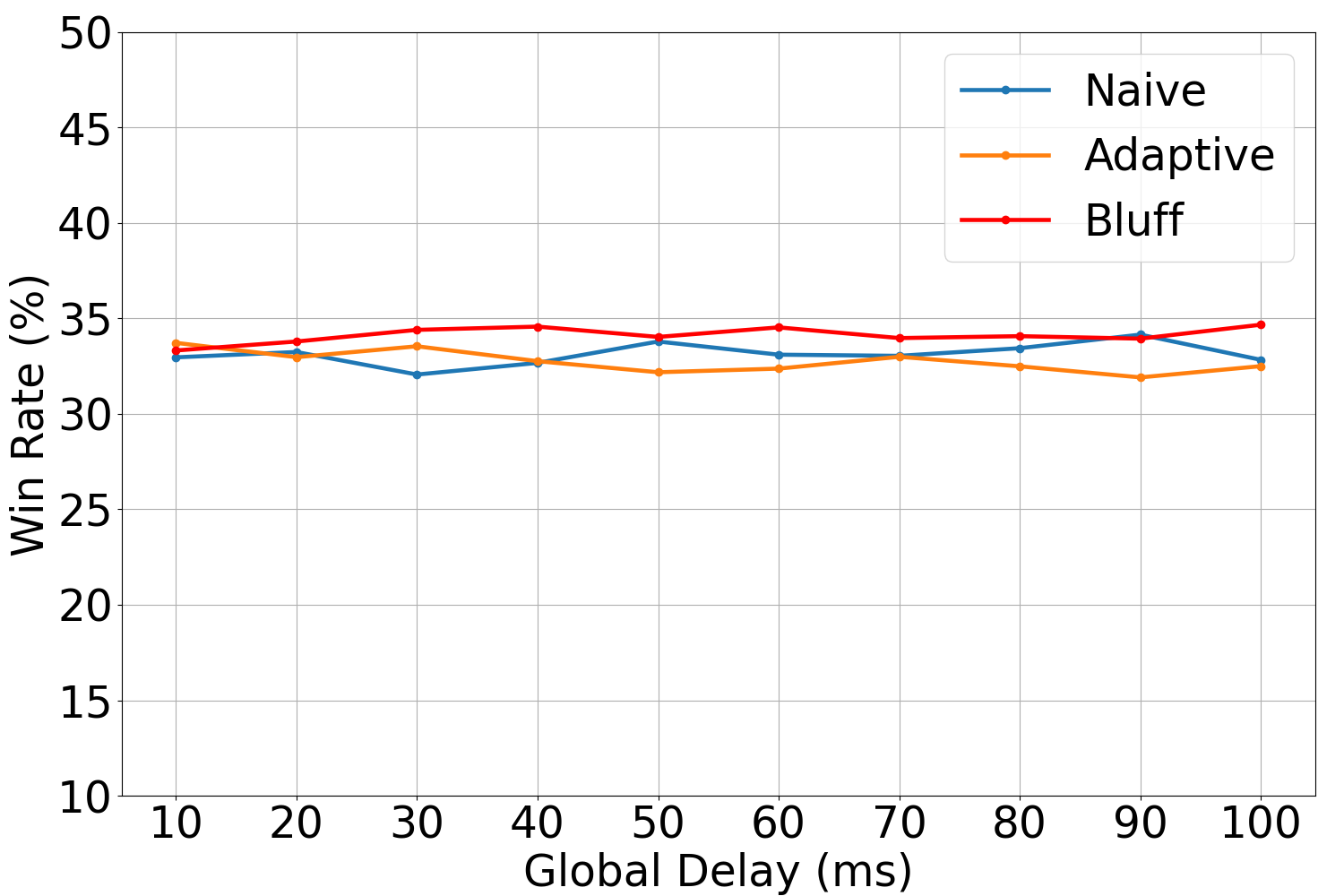}}
\caption{Win rate of strategies in Profile 1 (left) and Profile 2 (right) under varying global delays.}
\label{fig:winrate}
\vspace{-0.25cm}\end{figure}

\noindent {\bf Global delay.} 
Starting with strategy performance, Fig.~\ref{fig:winrate} illustrates the win rate performance of each strategy 
in both profiles under varying global delays from 10ms to 100ms. From the left plot for Profile 1, we observe 
that the win rate of adaptive players decreases by an average of $1.45\% \pm 0.68\%$ for every 10ms increase in global delay. This trend can be attributed to the reactive nature of the adaptive strategy in conjunction with the signal simultaneity. Specifically, while naive players directly update their bids, adaptive players are constrained by the delay period. They must wait until other players' bids are made eligible before 
responding, while other players may update their bids again during this period. This delay, coupled with the simultaneous 
signal updates, 
adversely impacts the effectiveness of the adaptive strategy for higher 
global delays. 

In Profile 2 (right plot of Fig.~\ref{fig:winrate}), the win rate performance of all three strategies remarkably remains the same around $33.3\%$. This can be largely explained by the bidding behavior of bluff players who conceal their actual valuation with bluff bids till the last moment of the auction. This behavior forces adaptive players to bid like naive players, leading to a scenario in which all players, regardless of the initial strategy, bid naively by the end of the auction. Since the revelation of bluff players inherently accounts for latency, this outcome occurs across all levels of global delay.

\begin{figure}[t]
\centering
{\includegraphics[width=1\linewidth]{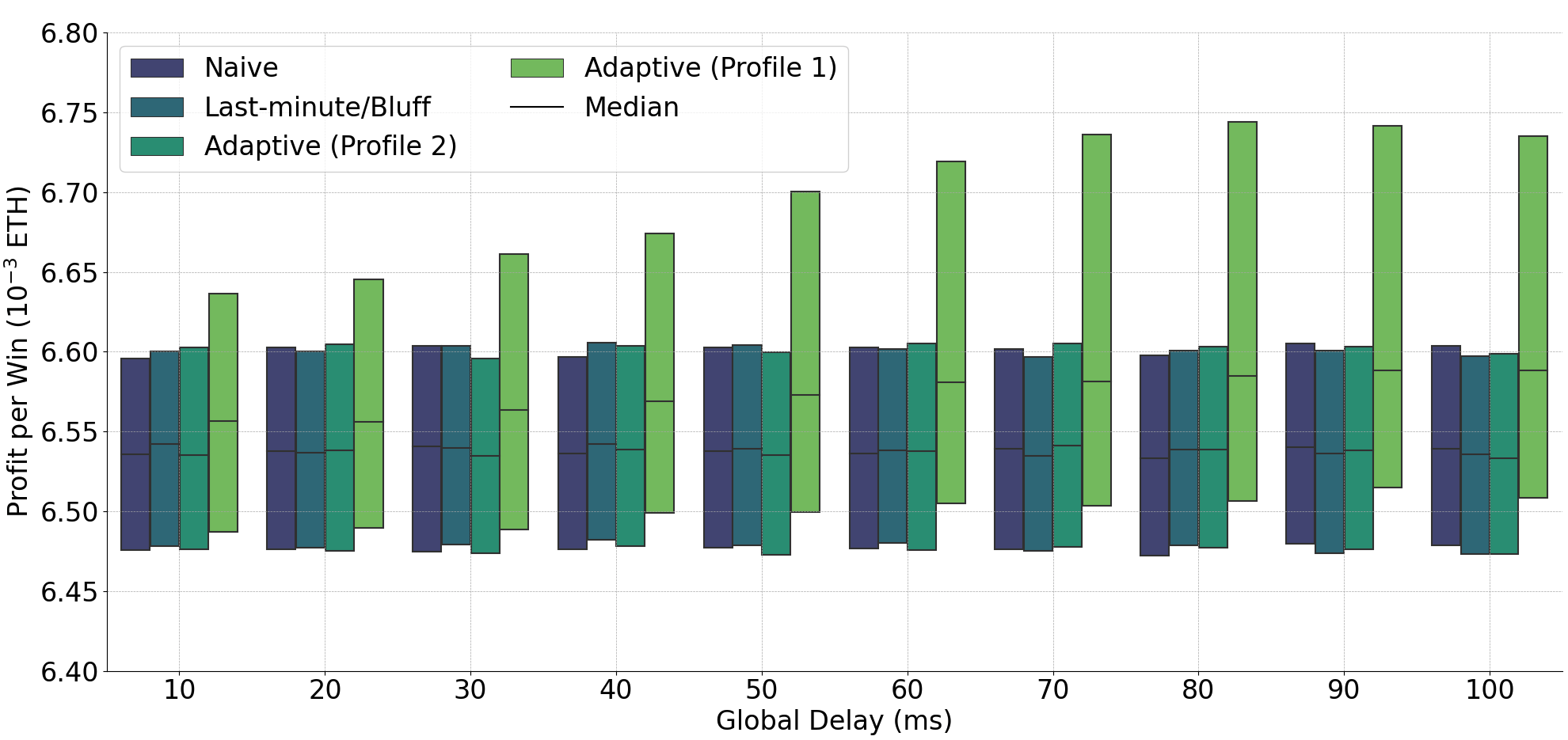}}
\caption{Profit-per-win distributions of all strategies for varying global delays. The lines in the box plots depict the 1st, 2nd (median), and 3rd quartiles of the data, respectively.}
\label{fig:ppw_global}
\vspace{-0.2cm}\end{figure}

Fig.~\ref{fig:ppw_global} shows the distribution of profit per win for each strategy 
under varying global delays. Notably, the profits per win for the naive, last-minute, bluff, and especially,  adaptive players in Profile 2 remain consistent regardless of the global delays. This uniformity arises from all players' employing naive-like bidding unaffected by latency. 

By contrast, in Profile 1, the median profit per win for adaptive players is initially about $0.73\%$ higher than others and increases by $0.11\% \pm 0.05\%$ for every 10ms rise in global delay. The absence of bluff players in Profile 1 allows adaptive players to fully exercise their ``adaptivity''. 
Their reactive mechanism allows them to win the auction by bidding slightly over the highest existing bid, which potentially leads to greater profits. 
Furthermore, under higher delays, adaptive players have more time to potentially access more valuable EOF before their next bid. They can still win by marginally exceeding the highest bid, thereby enhancing profitability, but also contributing to a greater variance in profit per win. 


\begin{figure}[t]
\centering
{\includegraphics[width=0.7\linewidth]{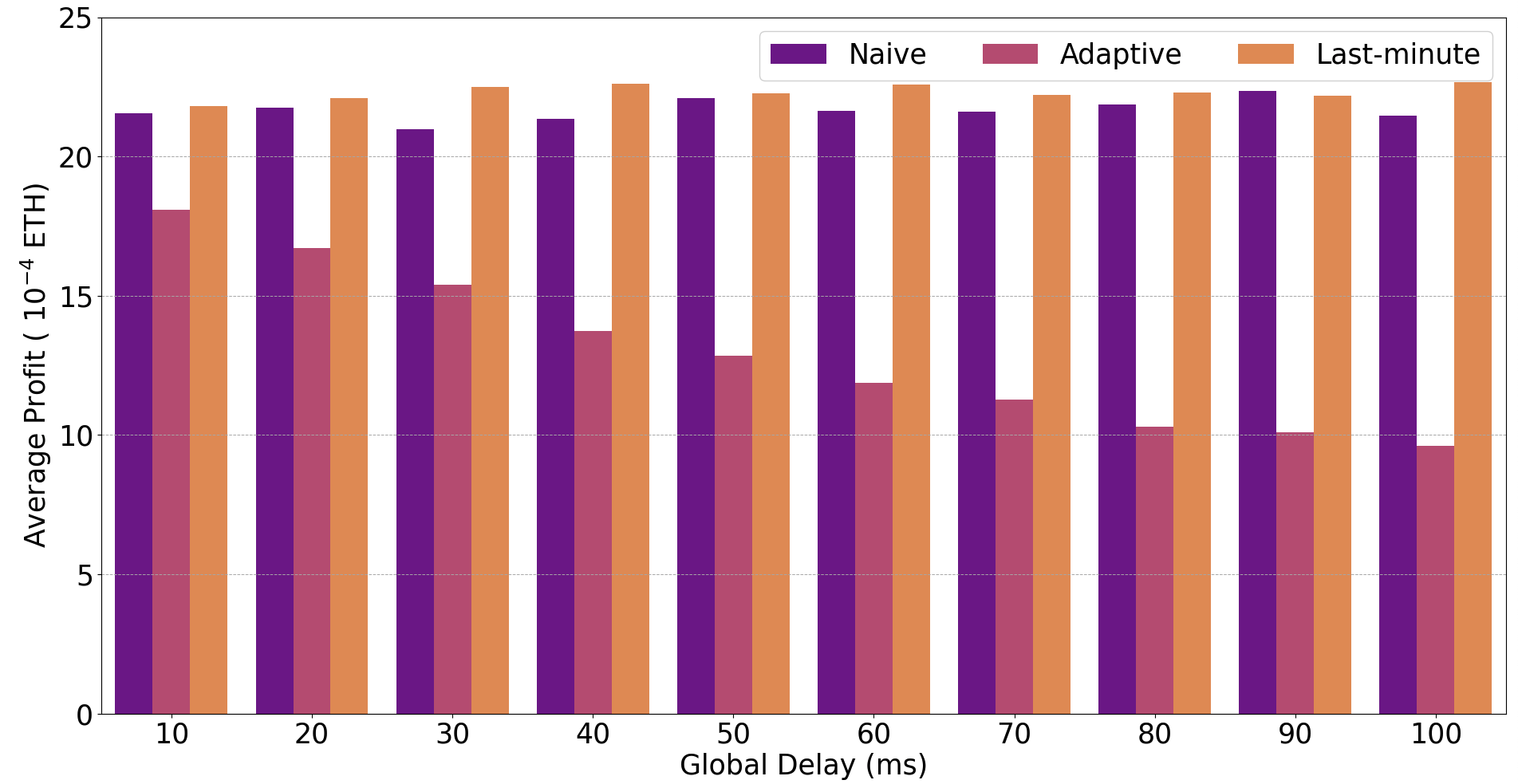}}
\caption{Average profits (Profile 1) under varying global delays.}
\label{fig:avg_p1}
\vspace{-0.2cm}\end{figure}

Nevertheless, while adaptive players in Profile 1 achieve a higher profit per win with increased global delays, their win rate drops significantly. These two conflicting effects contribute to an average decrease of $9.4\times 10^{-5}$ ETH in average profit per 10ms delay increase (Fig.~\ref{fig:avg_p1}), implying that higher global delays harm adaptive players after all. 
In Profile 2, as all players eventually turn naive, average profits across all strategies level out at approximately $2.2\times 10^{-3}$ ETH.

\begin{figure}[t]
\centering
{\includegraphics[width=0.49\linewidth]{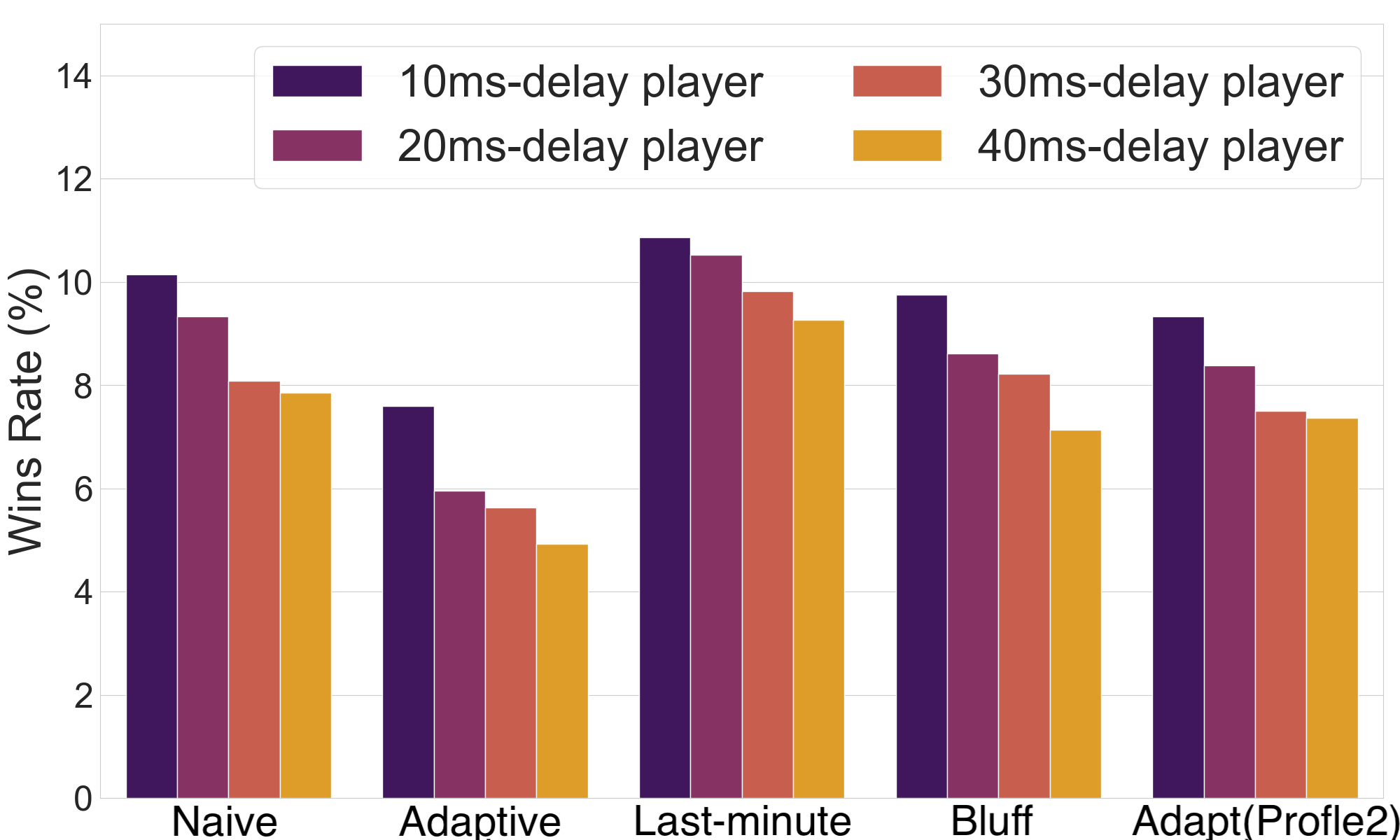}}
{\includegraphics[width=0.49\linewidth]{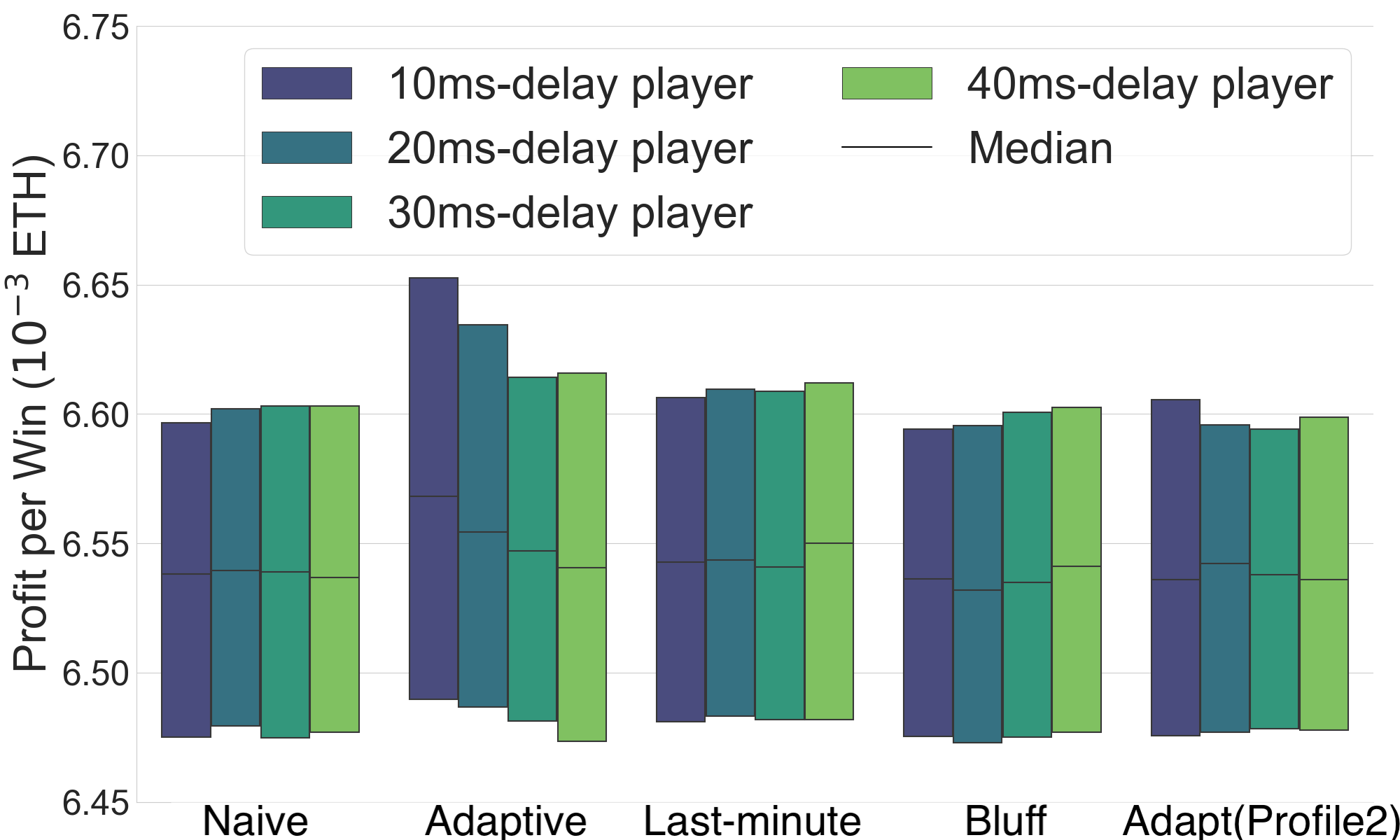}}
\caption{Win rate and profit per win performance of each player in both profiles (global delay = 10ms).}
\label{fig:indidelay}
\vspace{-0.2cm}\end{figure}

\noindent {\bf Individual delay.}
%
Fig.~\ref{fig:indidelay} shows the performance of each player in their respective strategy group under a global delay of 10ms. 
The left plot indicates that for all players in both profiles\footnote{We omit naive players in Profile 2 as they do not add new complexities.}, a 10ms decrease in individual delay can contribute to an approximately $0.68\%$ advantage in win rate, as players can reach the relay sooner. 
This is particularly beneficial when MEV opportunities emerge late in the auction. 

The rationale behind the increasing trend of adaptive players' profit per win under higher global delays also applies to the results in the right plot of Fig.~\ref{fig:indidelay}.
In particular, adaptive players with lower individual delays have an enhanced ability to include more MEV-profitable transactions in their block, as they can submit their bid later, 
resulting in a higher profit per win. 
Although this advantage 
does not apply 
to other naive-like players, such players should still aim to reduce 
their individual delay for 
higher win rates. 

\smallskip 
\noindent \underline{\it Impact of revealing time.}
Last-minute and bluff players 
reveal 
their valuations at different time points before the auction terminates. We delve into the impact of the revealing time, $\epsilon$, of these 
players on their performance and strategy effectiveness. To study the trade-off between revealing time and latency and how it affects strategy effectiveness, 
we standardize the individual delay at 10ms for all players across profiles and conduct simulations under a consistent global delay with a fixed 12-second auction interval.\par

\begin{figure}[t]
\centering
{\includegraphics[width=0.49\linewidth]{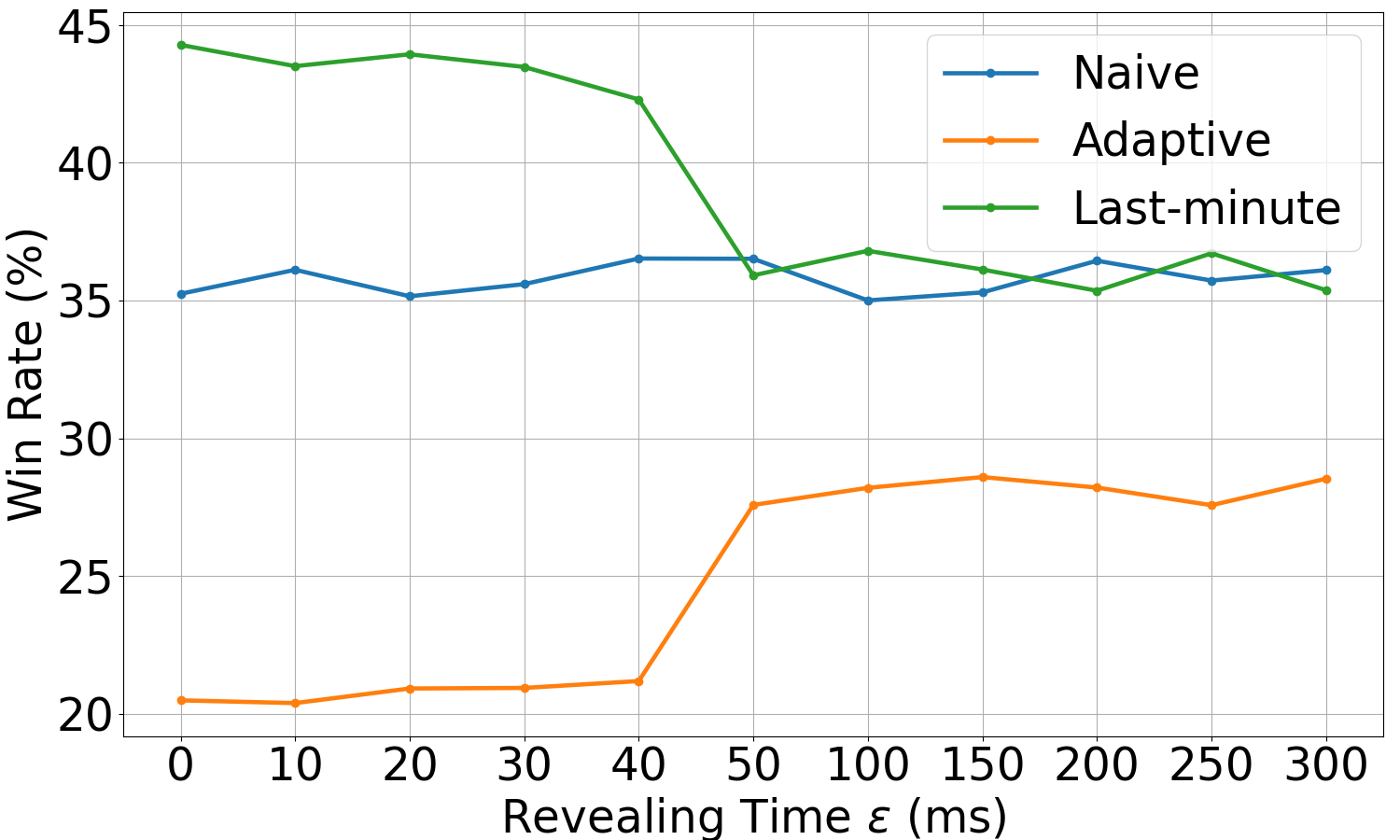}}
{\includegraphics[width=0.49\linewidth]{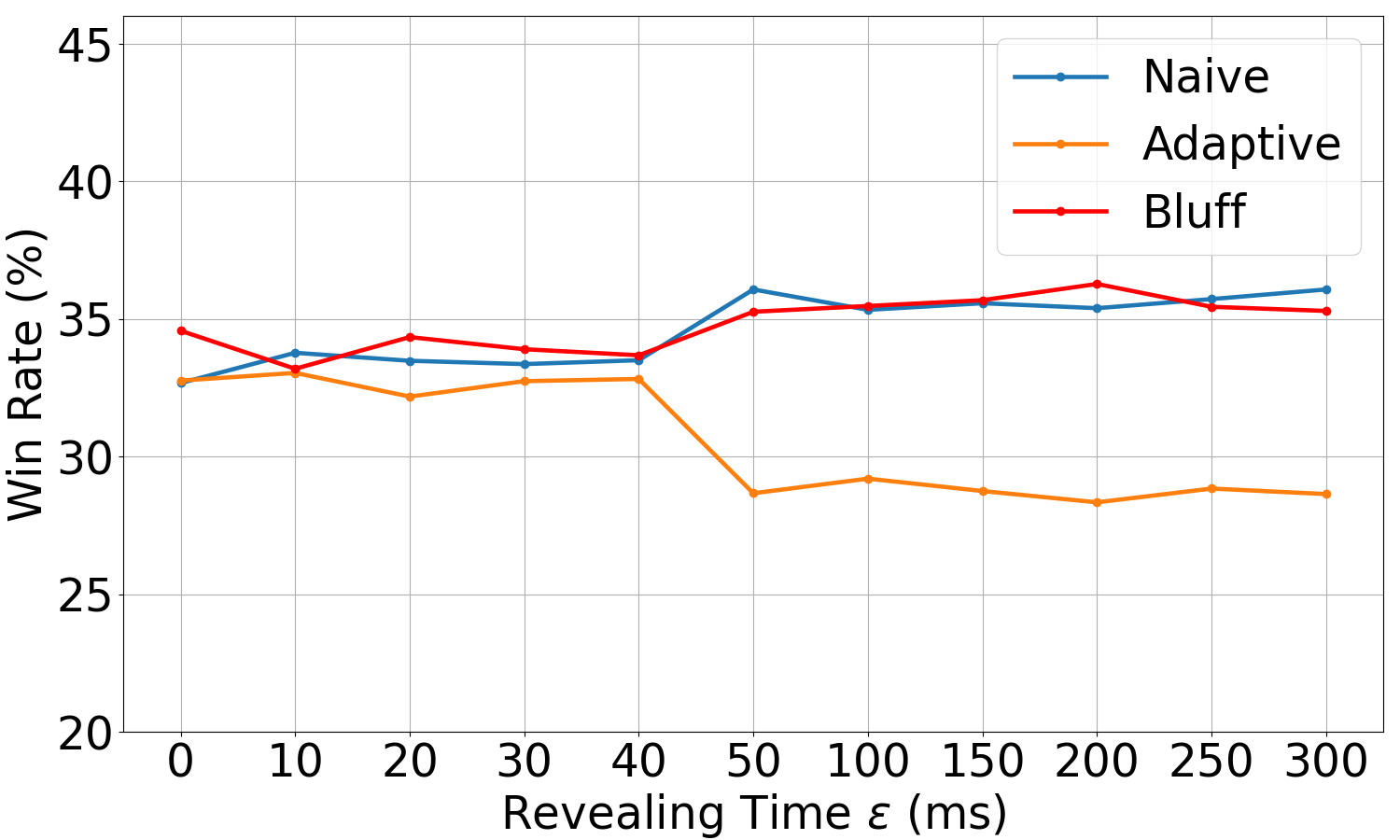}}
\caption{Win rate of strategy groups in Profile 1 (left) and Profile 2 (right) considering varying revealing times when global delay = 40ms.}
\label{fig:epsilon_winrate}
\vspace{-0.2cm}\end{figure}

Fig.~\ref{fig:epsilon_winrate} assesses the win rate performance of each strategy 
in both profiles, considering varying revealing times for last-minute and bluff strategy, with a constant 40ms global delay\footnote{We present 40ms, but the results are analogous across different values.}. In the left plot, we observe a significant advantage in the win rate for last-minute players compared to naive players in Profile 1 for values of $\epsilon<50$ms. As can be also seen from the left plot of Fig.~\ref{fig:winrate}, this advantage is robust across various global delays, taking values in the range $7.45\% \pm 1.75\%$, and can be attributed to the revelation of last-minute players' valuations at the final time step of the auction. Against typical naive players, adaptive players have the capacity to react and update their bids in response to emerging MEV opportunities. However, the strategic behavior of last-minute players prevents adaptive players from reacting effectively before the auction terminates. This offers last-minute players a strategic edge over adaptive players, enabling them to outperform typical naive players. \par
From Fig.~\ref{fig:epsilon_winrate}, however, it becomes apparent that the advantage of last-minute players diminishes when they reveal their valuation 50ms or more before the auction ends. This observation is explainable by considering the latency faced by adaptive players. The total latency for 
adaptive players amounts to 50ms (i.e., 40ms global delay plus a 10ms individual delay). Thus, if last-minute players reveal their bids later than 50ms before the auction terminates (i.e., after 11.95s), 
the response of adaptive players cannot be accepted by the relay in time. 
In contrast, if last-minute players reveal their bid 
before 11.95s, adaptive players have sufficient time to submit a competitive bid before the auction closes.

In the right plot of Fig.~\ref{fig:epsilon_winrate} for Profile 2, a shift in performance among the three strategy groups is also observed at the same point. 
When the revealing time of bluff players is less than the total latency of adaptive players, adaptive players are forced to act like naive players throughout the effective bidding period. 
When the revealing time exceeds the total latency of adaptive players, 
adaptive players regain their ability to respond dynamically after bluff bids are canceled but still struggle with latency and signal simultaneity.

\begin{figure}[t]
\centering
{\includegraphics[width=0.49\linewidth]{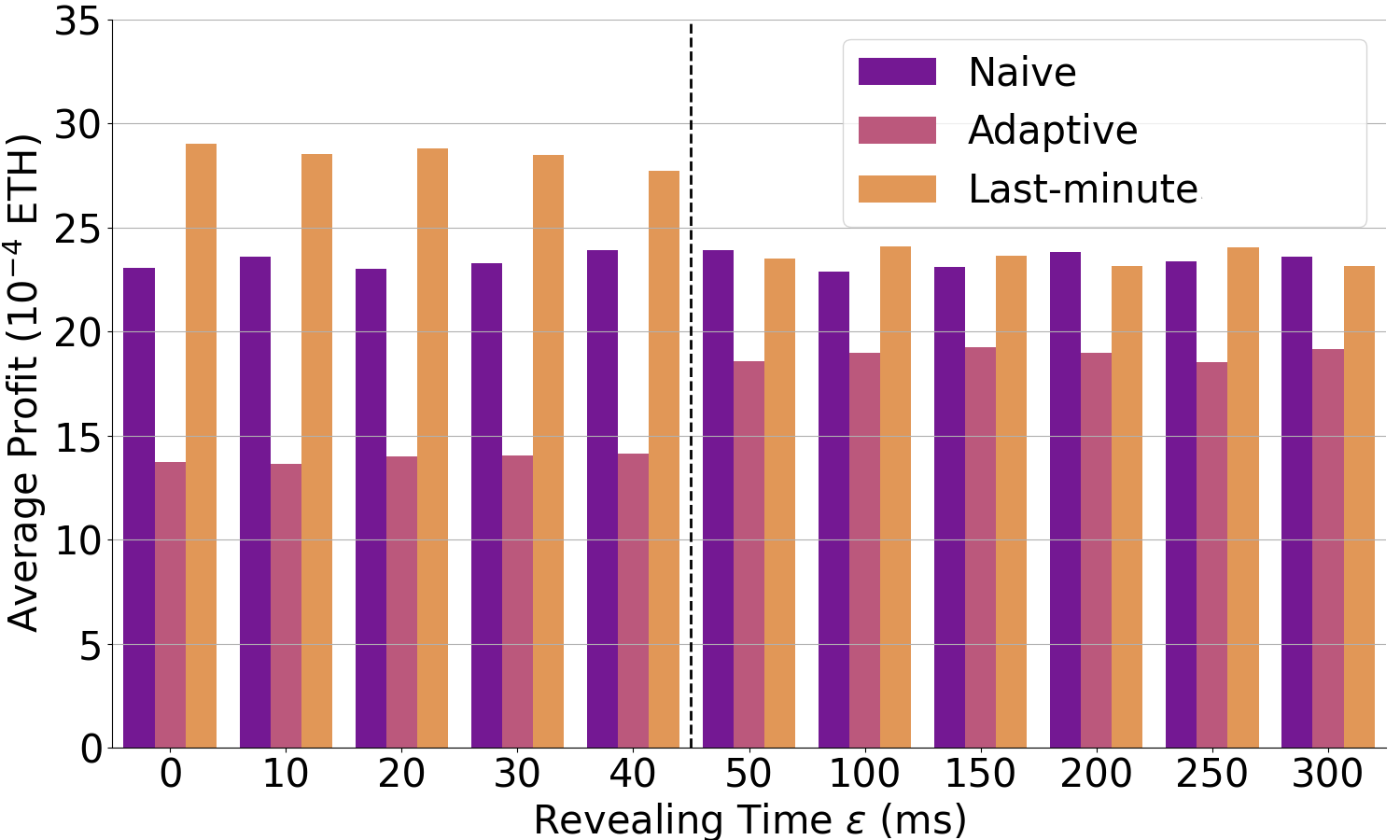}}
{\includegraphics[width=0.49\linewidth]{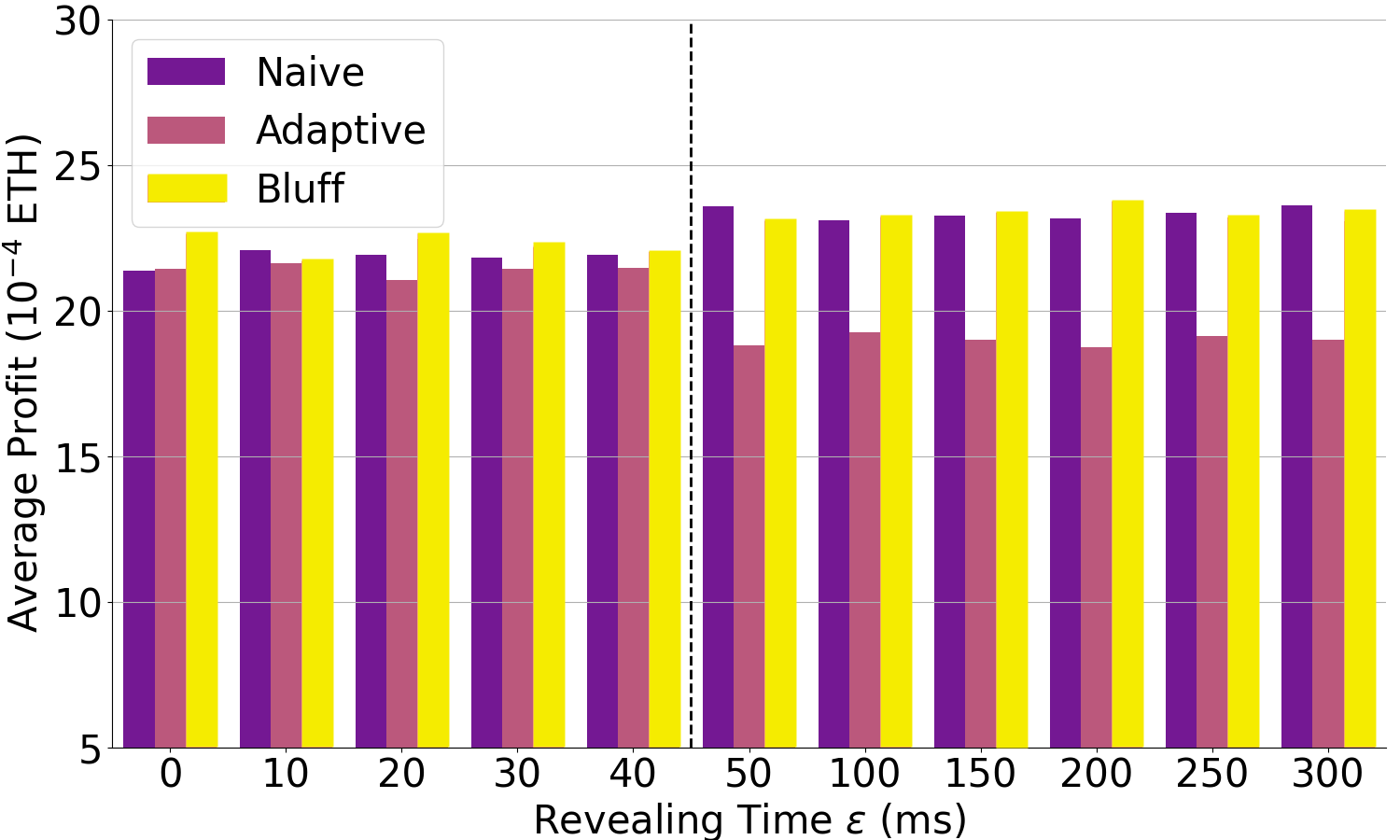}}
\caption{Average profit of strategy groups in Profile 1 (left) and Profile 2 (right) under varying revealing times when global delay = 40ms.}
\label{fig:epsilon_profit}
\vspace{-0.25cm}\end{figure}

We observe in Fig.~\ref{fig:epsilon_winrate} that when adaptive players can respond dynamically after the revelation, their win rate converges to around $28\%$, while the win rate of other naive-like players converges to around $36\%$. These findings remain valid vis-\`a-vis average profit, see Fig.~\ref{fig:epsilon_profit}.

Based on the analysis above, in 
fixed interval scenarios, the strategic timing advantage makes the last-minute strategy more effective. 
By revealing their bids at the very end of the auction, last-minute players constrain adaptive players' response time, thereby reducing their performance. While bluff players compel adaptive players to commit to their valuations, this does not significantly impede adaptive players from reacting effectively but pushes them to bid more aggressively. 

\begin{figure}[t]
\centering
{\includegraphics[width=0.8\linewidth]{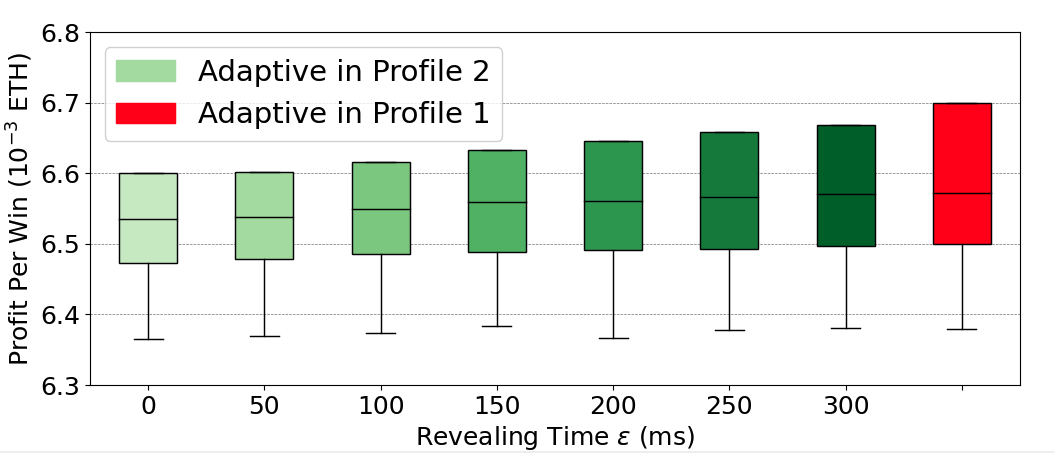}}
\caption{Adaptive players' profitability under revealing times of bluff players.}
\label{fig:ppp_adapt}
\vspace{-0.2cm}\end{figure}

Finally, Fig.~\ref{fig:ppp_adapt} showcases the impact of bluff players' revealing time on adaptive players' profit per win by comparing the profit per win of adaptive players under different bluff revealing times (green boxes) including scenarios without bluff players (red box). The profit per win of adaptive players in Profile 2 and the earliness of the bluff players’ cancellations exhibit a positive correlation, with an average increase of $5.9\times 10^{-6}$ ETH in the median for every 50ms increase in revealing time. 
However, even if bluff bids are canceled earlier, the profit per win of adaptive players in Profile 2 remains lower compared to those in Profile 1, highlighting the significant impact of bluff strategy on adaptive strategy's profitability.

\smallskip \noindent \underline{\it Impact of auction termination time.} 
In practice, the MEV-Boost auction 
does not always end precisely at $T = 12$.
Uncertain auction intervals can have a significant impact on player performance, particularly for last-minute and bluff players, who may fail to reveal their valuation on time.

To study this effect, we simulate scenarios with a random auction interval by adjusting the value of $\sigma$. 
%
To control for latency factors, we standardize the individual delay at 10ms for all players across both profiles, while maintaining a global delay of 10ms. 
In addition, we include two distinct scenarios, one for zero and one for non-zero revealing times to assess how players' different strategic decisions influence strategy effectiveness under a random-interval auction setup. To ensure that bluff bids are generally higher than the total signal, their values are randomly drawn from a uniform distribution ranging between 0.3 and 0.4 ETH. We choose to study this range, but the results are analogous to any higher value.

\noindent {\bf Revealing time \texorpdfstring{$\epsilon = 0$}{epsilon = 0}.}
For last-minute and bluff players, revealing their valuation at the final expected moment of the auction is a strategic move, as they can benefit from limiting adaptive players' reaction time. However, 
due to the inherent uncertainty in the auction's actual end time, revealing exactly at 12 seconds means that these players have only a 50\% chance of successfully revealing their bids before the auction closes. For last-minute players, failing to reveal their valuation before the auction terminates means missing the chance of winning because they hold back their bids. Conversely, bluff players will end up winning the auction at negative profits, if they fail to cancel their bluff bid before the auction terminates. 

\begin{figure}[t]
\centering
{\includegraphics[width=0.49\linewidth]{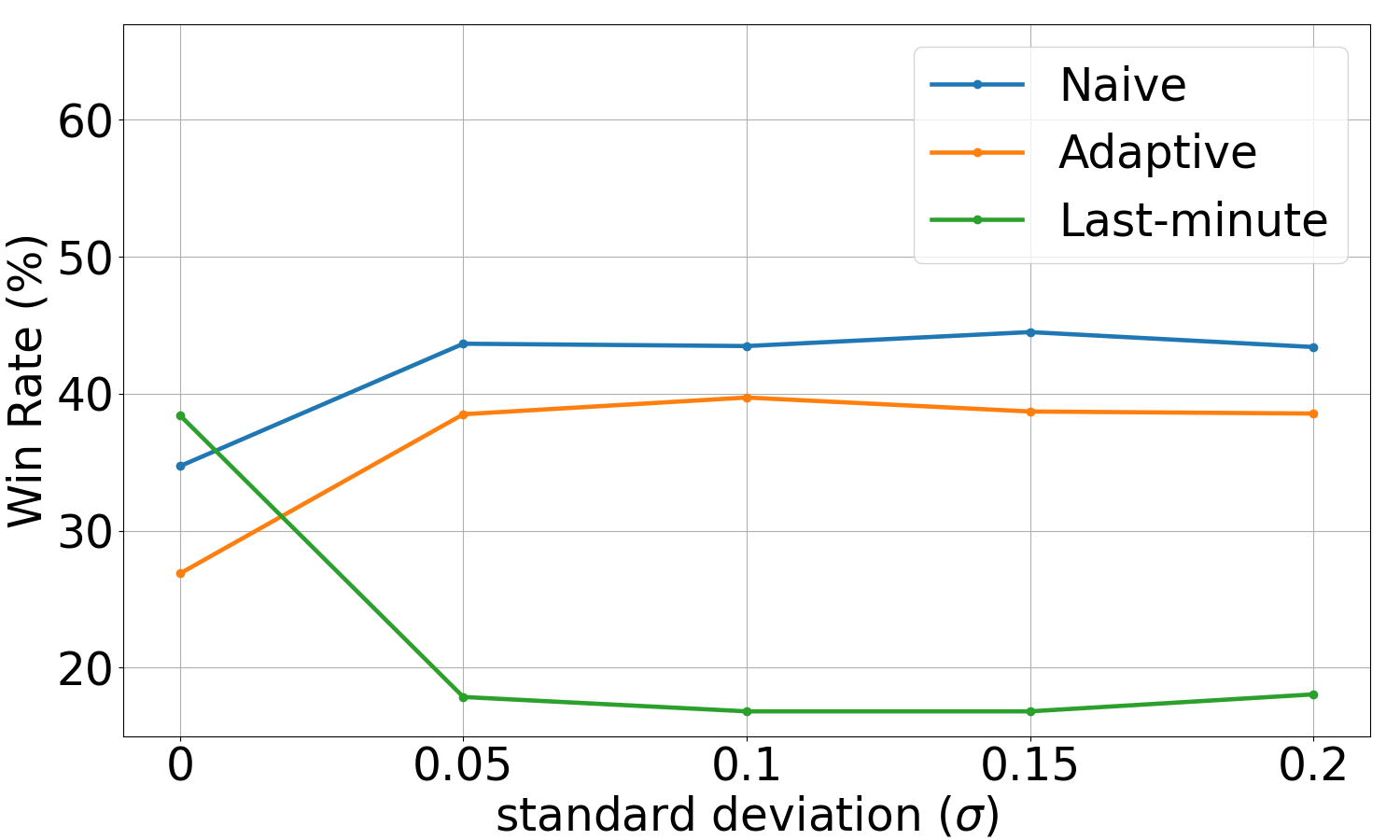}}
{\includegraphics[width=0.49\linewidth]{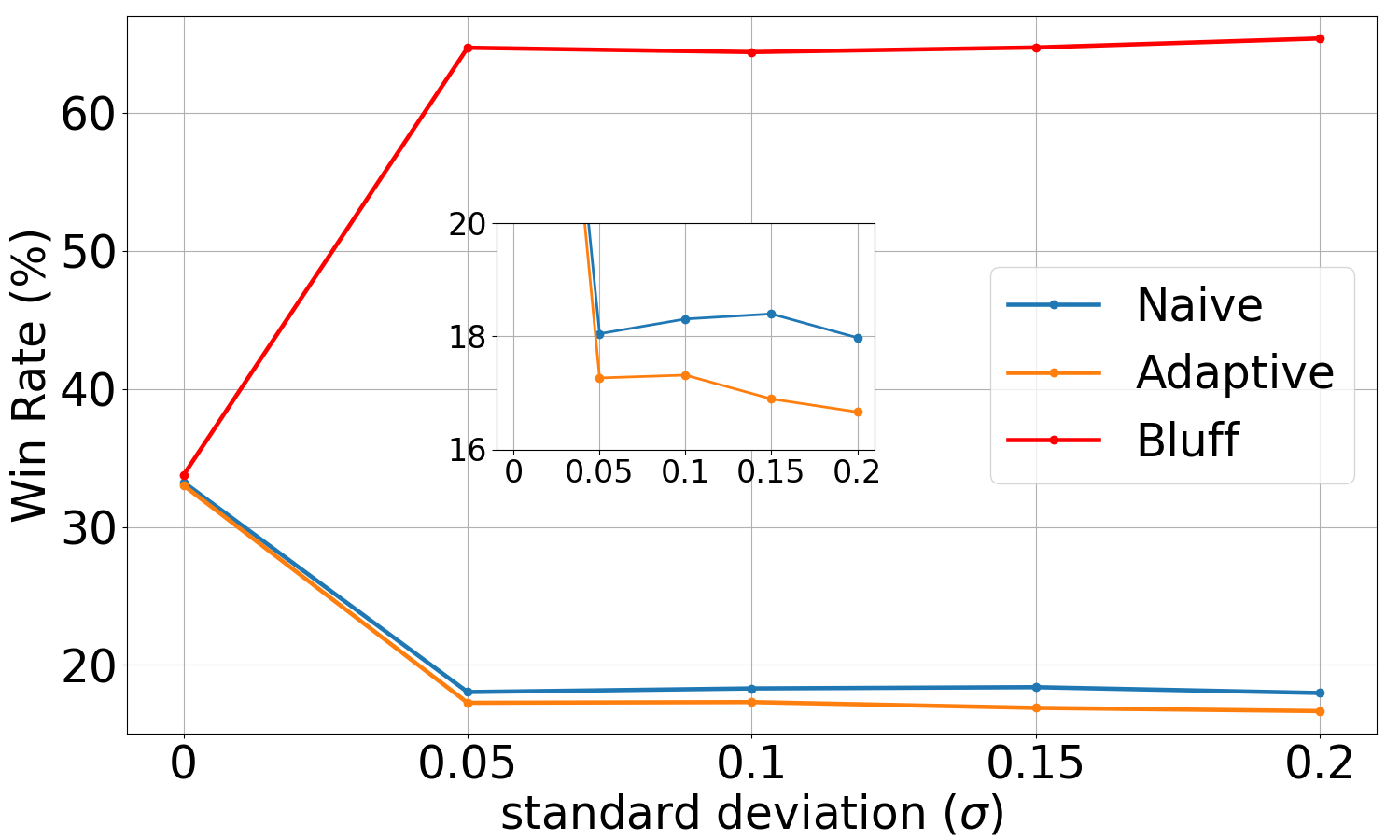}}
{\includegraphics[width=0.49\linewidth]{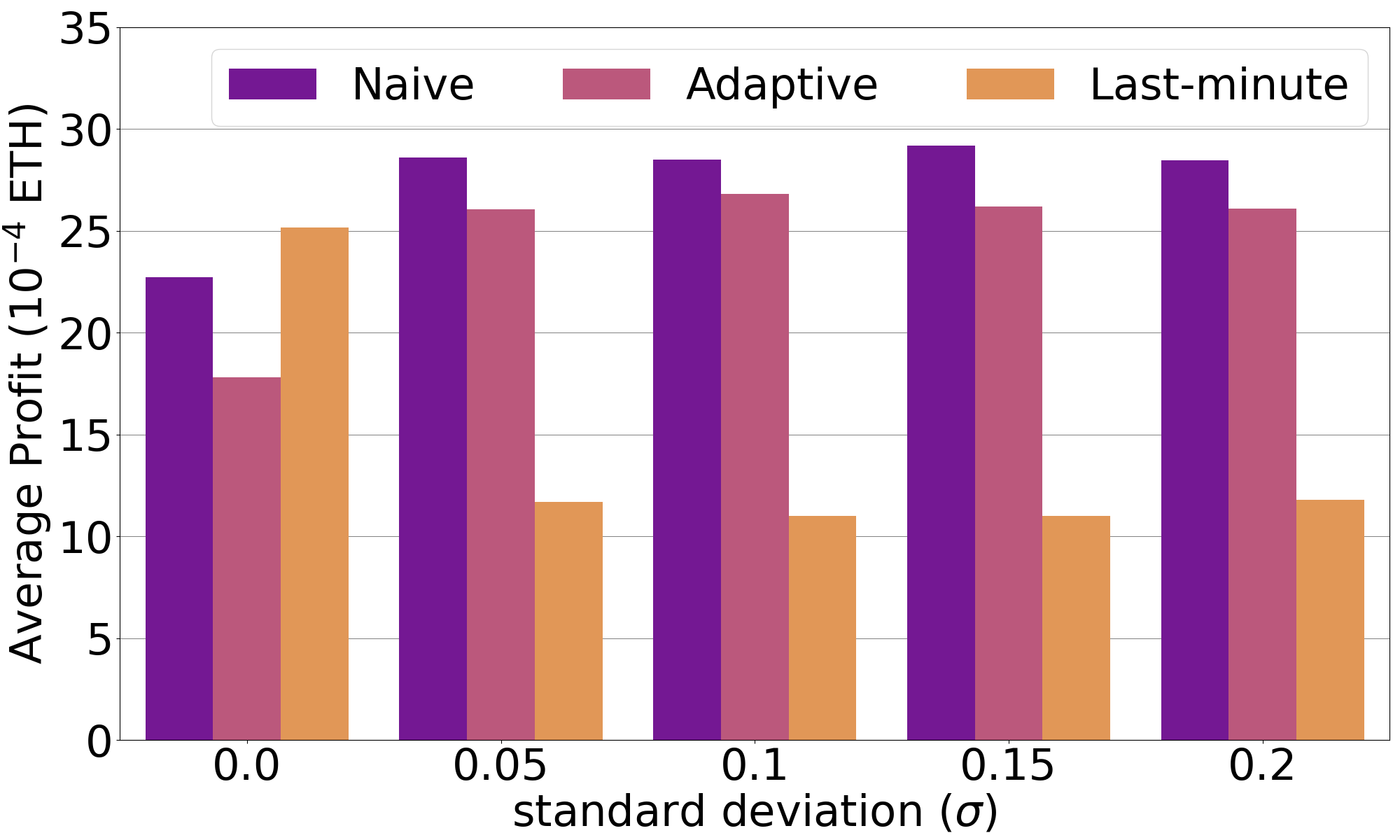}}
{\includegraphics[width=0.49\linewidth]{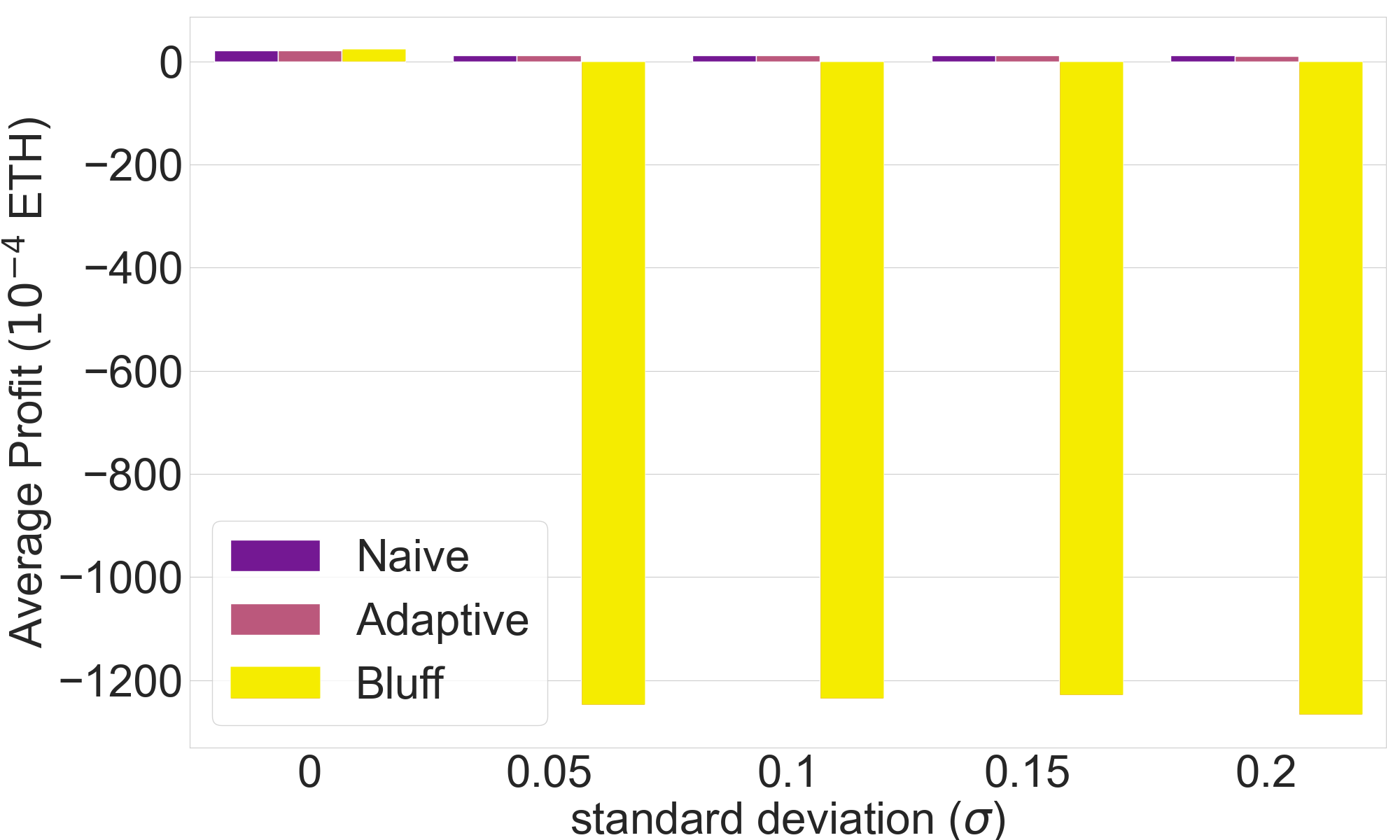}}
\caption{Overall performance of strategy groups in Profile 1 (left) and Profile 2 (right) under varying standard deviations when revealing time $\epsilon$ = 0.}
\label{fig:std_0}
\vspace{-0.2cm}\end{figure}

Fig.~\ref{fig:std_0} showcases the win rates and average profits of each strategy 
in both profiles when $\epsilon = 0$. In the left two plots for Profile 1, last-minute players initially excel when $\sigma = 0$ by revealing at the last moment of a fixed-interval auction. 
But as the auction interval starts to vary around 12 seconds, their performance drops and converges to an average profit of $1.1\times10^{-3}$ ETH with an $18.3\%$ win rate under varying $\sigma$ values ($\sigma\neq 0$), due to a $50\%$ likelihood of revealing their bids too late and losing the auction. Similarly, in the right two plots for Profile 2, 
we see bluff players 
canceling their bluff bids too late half the time and winning the auction, 
contributing to a $65.1\%$ win rate on average, but a negative average profit of approximately $-0.12$ ETH. 
Naive and adaptive players exhibit similar performances, as bluff players force adaptive players to behave like naive players. However, 
adaptive players increasingly lag behind as the standard deviation grows, as is shown in the zoomed-in plot. This is attributed to a higher chance of the auction terminating at a relatively later time after 12 seconds, providing adaptive players with more time to be truly ``adaptive'' after bluff bids are canceled at 12 seconds. Consequently, adaptive players exhibit relatively lower win rates in such scenarios.

\begin{figure}[t]
\centering
{\includegraphics[width=0.49\linewidth]{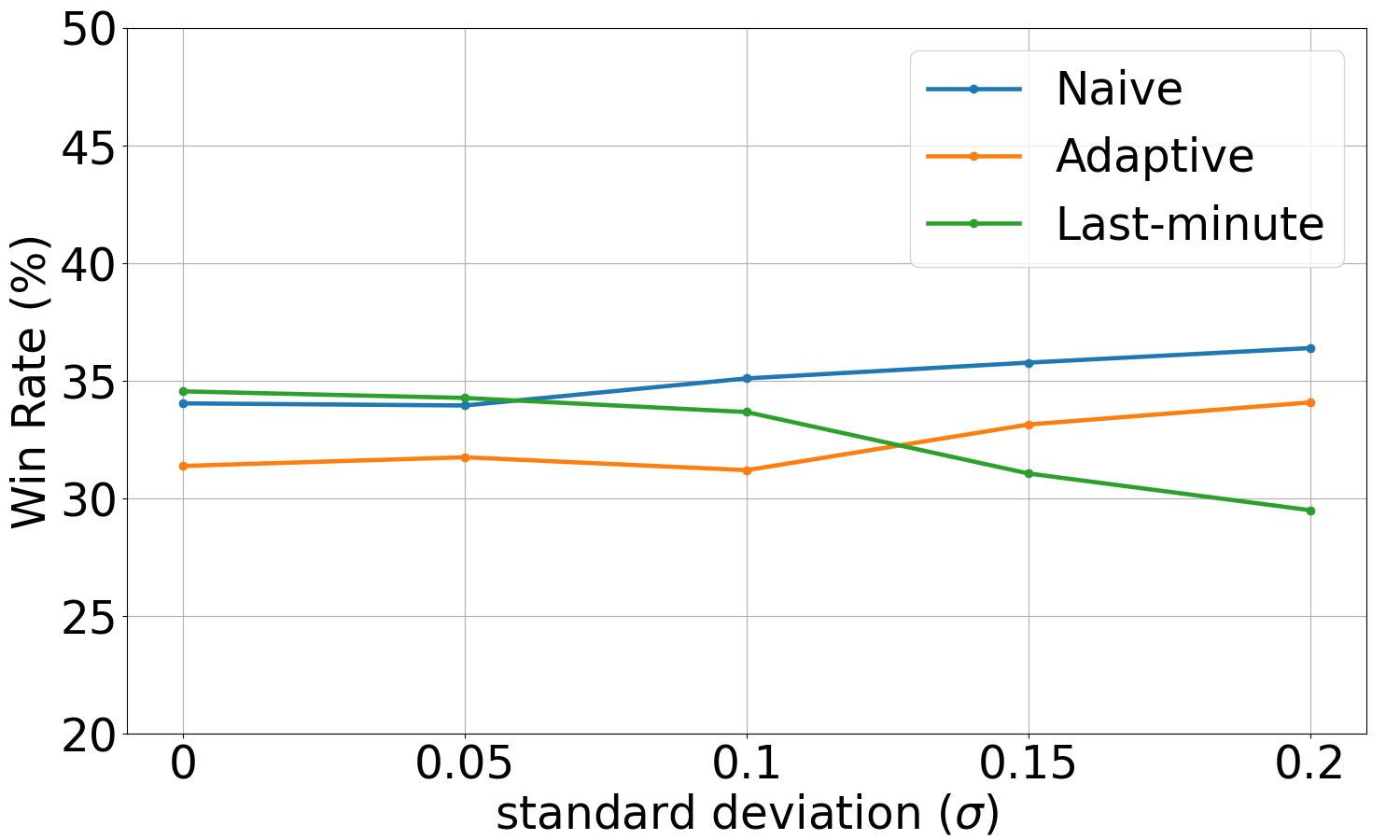}}
{\includegraphics[width=0.49\linewidth]{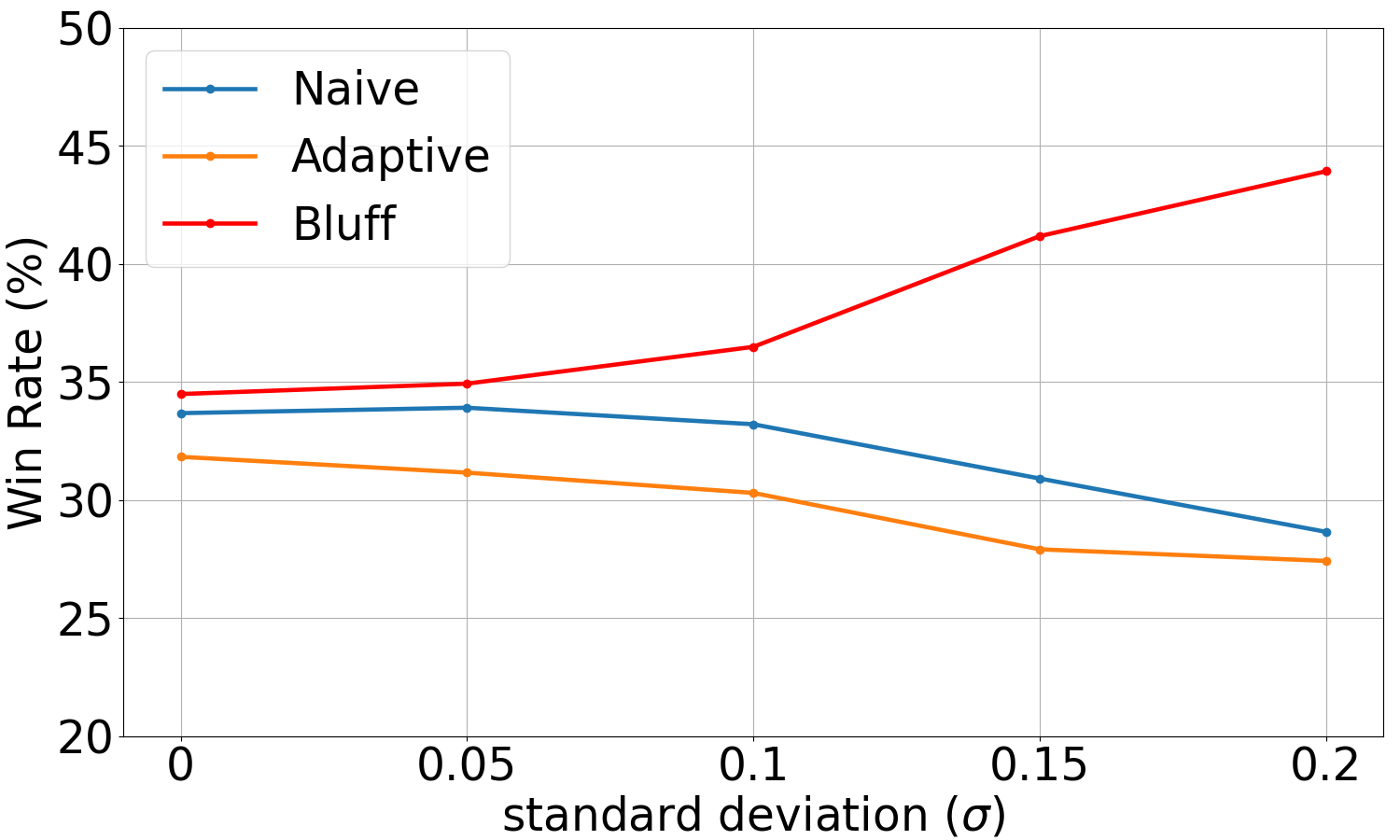}}
{\includegraphics[width=0.49\linewidth]{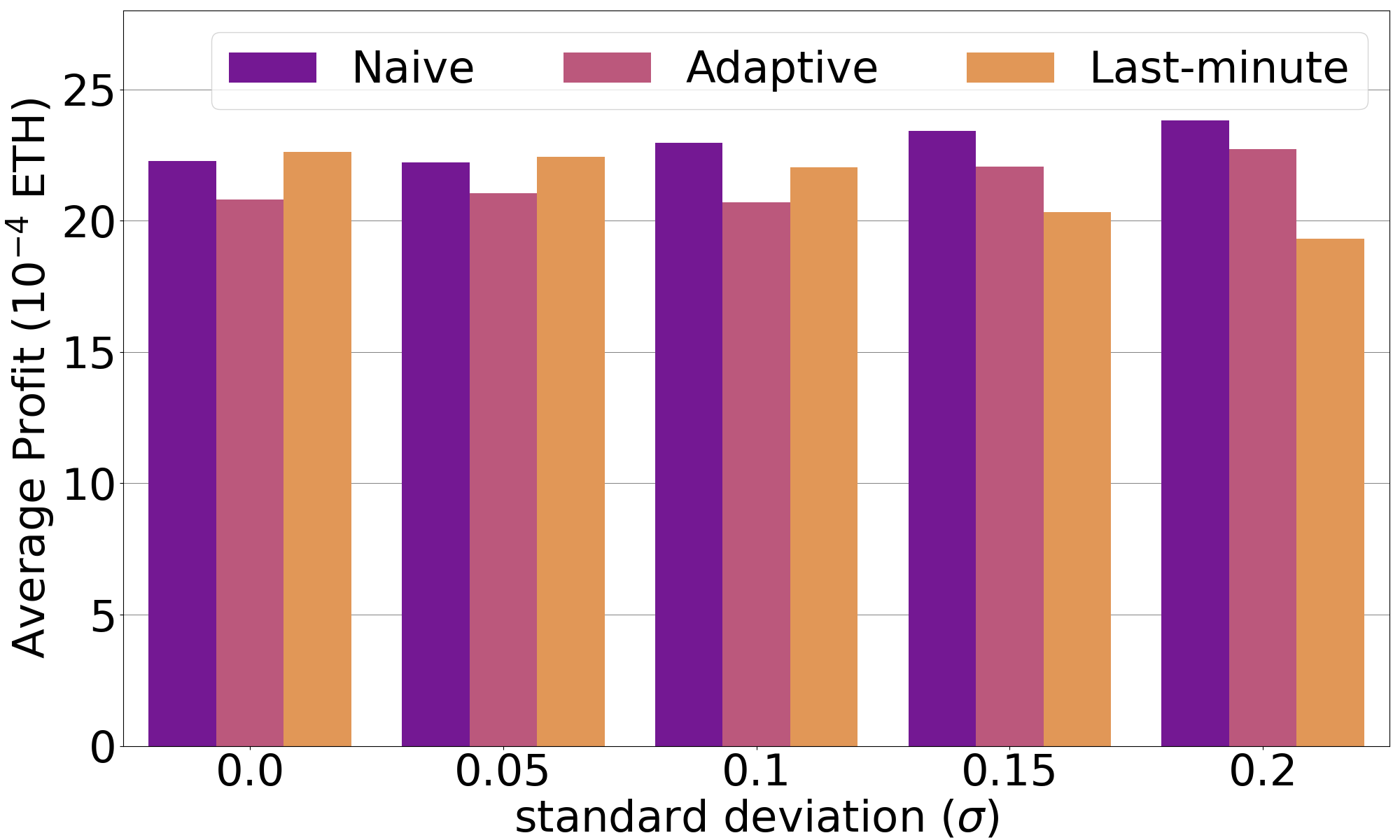}}
{\includegraphics[width=0.49\linewidth]{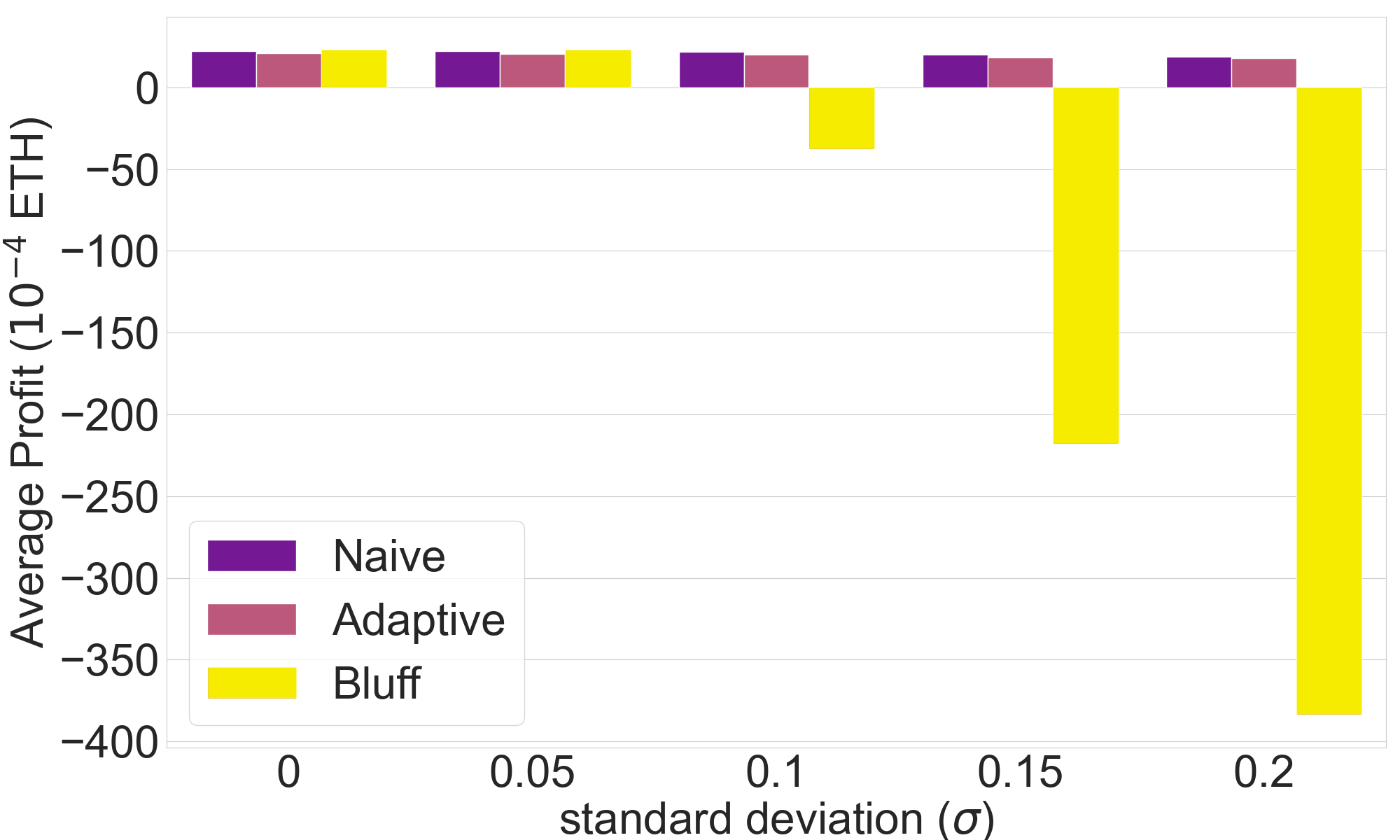}}
\caption{Overall performance of strategy groups in Profile 1 (left) and Profile 2 (right) under varying standard deviations when revealing time $\epsilon = 0.2$.}
\label{fig:std_02}
\vspace{-0.2cm}\end{figure}

\noindent {\bf Revealing time \texorpdfstring{$\epsilon > 0$}{epsilon > 0}.} 
Last-minute and bluff players can reveal earlier to mitigate the uncertainty associated with the auction interval. We choose to study $\epsilon = 0.2$, but the results are analogous across different values. From Fig.~\ref{fig:std_02}, we see that when last-minute players reveal their bids at 11.8 seconds into the auction and then become naive-like players, they exhibit performance levels comparable to naive players 
with low auction interval deviation. 
However, as the deviation degree increases, the performance of last-minute players starts to deteriorate. This decline can be attributed to the escalating probability of an auction termination before 11.8 seconds, eventually contributing to better performance of naive and adaptive players. Similarly, bluff players are as competitive as naive players when the deviation is relatively low. Their positive average profit shows that they successfully cancel their bluff bids before the auction terminates. As the degree of deviation increases, bluff players start to win with bluff bids, which, again, leads to a negative profit.

This analysis shows last-minute players' strategic advantage gained from the late revelation in fixed-interval auctions diminishes as uncertainty in the termination of auction interval increases.
The situation is more challenging for bluff players, who risk winning the auction but incurring negative profits due to the timing of their bid cancellation. Even when these players reveal their bids slightly earlier, their performance is still susceptible to varying degrees of auction interval deviation. This adjustment also results in a loss of their strategic advantage over other players, underscoring the critical influence of auction interval deviation on strategy effectiveness.

\section{Discussion: Performance of Strategies}
\label{sec:discuss}
Based on the above analysis, we can draw the following conclusions on the impact of MEV-Boost auction factors on the effectiveness and the distinctive characteristics of the commonly-used MEV-Boost strategies that we simulated.\par\smallskip
\emph{A. MEV-Boost auction factors:} Our findings underscore the impact of latency in MEV-Boost auctions and the need for players to strategically consider these elements to enhance their performance. 
In our model, the global delay represents the bid processing time imposed uniformly by relays on all builders. It can serve as a key determinant in differentiating outcomes under optimistic versus non-optimistic relay scenarios in MEV-Boost auctions. Unlike non-optimistic relays that verify the block content immediately after receiving a block, optimistic relays \cite{optimisitcrelay} assume blocks are valid in the short term and defer verification to a later time\footnote{A builder must provide collateral when submitting blocks to optimistic relays. If a builder submits an invalid block that wins the auction and is proposed, the block will be discarded and the proposer who missed the slot will be refunded the amount of the winning bid \cite{optimisitcrelay}.}, leading to faster bid acceptance. Our findings suggest the adaptive strategy is not as effective as other strategies in non-optimistic relay scenarios.\par
The lower win rates and average profits under higher individual delays, particularly for adaptive players who already need to compensate for their additional reaction time, also suggest that builders should improve their network connection to the relay, possibly considering co-location to the relay server. Furthermore, since the effect of global delay and individual delay is equivalent in our model, individual delay also reflects a player's strategic choice between submitting bids to optimistic or non-optimistic relays. Our results suggest that all strategies benefit from shorter delays, and support the wide adoption of optimistic relays by builders.\par

Finally, the differences in performance between Profiles 1 and 2 also indicate the indirect impact of bluff players, particularly on adaptive players. Forcing adaptive players to reveal their valuations may reduce their profits per win, but may also come with a disadvantage for bluff players. In particular, bluff players make adaptive players immune to latency and more competitive in high-latency scenarios.

\emph{B. MEV-Boost auction strategies:} Based on our analysis, we can 
also summarise the distinctive characteristics of each strategy as follows. Naive players, although not bidding strategically, remain highly competitive across various auction setups due to their aggressive, consistent, valuation-driven bidding, which is not influenced by the auction setup.

Compared to naive-like players, adaptive players struggle with their win rate because of their reactive mechanism and the universal signal simultaneity. While they react on the highest bid, naive-like players might already submit an updated bid, making adaptive players' reactions outdated. This can be even worse when the latency becomes higher, 
as their reaction becomes slower. However, their reactive mechanism allows them to optimize their profit per win, especially 
in high-latency environments, but such an advantage vanishes with the existence of bluff players in the auction. \par

The last-minute strategy is very effective in fixed-interval auctions, as last-minute players can reveal their valuation at the final moment 
of the auction, leaving no response time for other players. However, in auctions with uncertain intervals, they can fail to reveal their bid before the auction closes and lose the chance of winning. When they choose to reveal their bids earlier to account for the uncertainty of the interval, they encounter the cost of losing the edge against other players. Last-minute players are not affected by latency, as their revealing action takes latency into account.\par
Finally, bluff players 
are primarily focused on influencing other players rather than winning. Bluff players force adaptive players to become more competitive, thereby losing their own edge and under uncertain auction intervals, they even put themselves at the risk of winning with negative profits.

\section{Auction Outcome}
\label{sec:outcome}

Next, we evaluate the overall auction outcome, with a primary focus on the proposer's perspective and 
the winning bid value. 
Our findings suggest that the global delay is a key factor shaping the auction outcome.\par

From the proposer's standpoint, the ideal scenario, apart from the exceptional case where a bluff bid wins, involves all players employing the naive strategy with full EOF access. Therefore, our strategy profile exclusively consists of naive players with full EOF access, with other player-specific variables being symmetric among all players. To ensure robust results, we run the auction 10,000 times with 10 naive players for each global delay setting, ranging from 10ms to 200ms.\par

\begin{figure}[t]
\centering
{\includegraphics[width=0.7\linewidth]{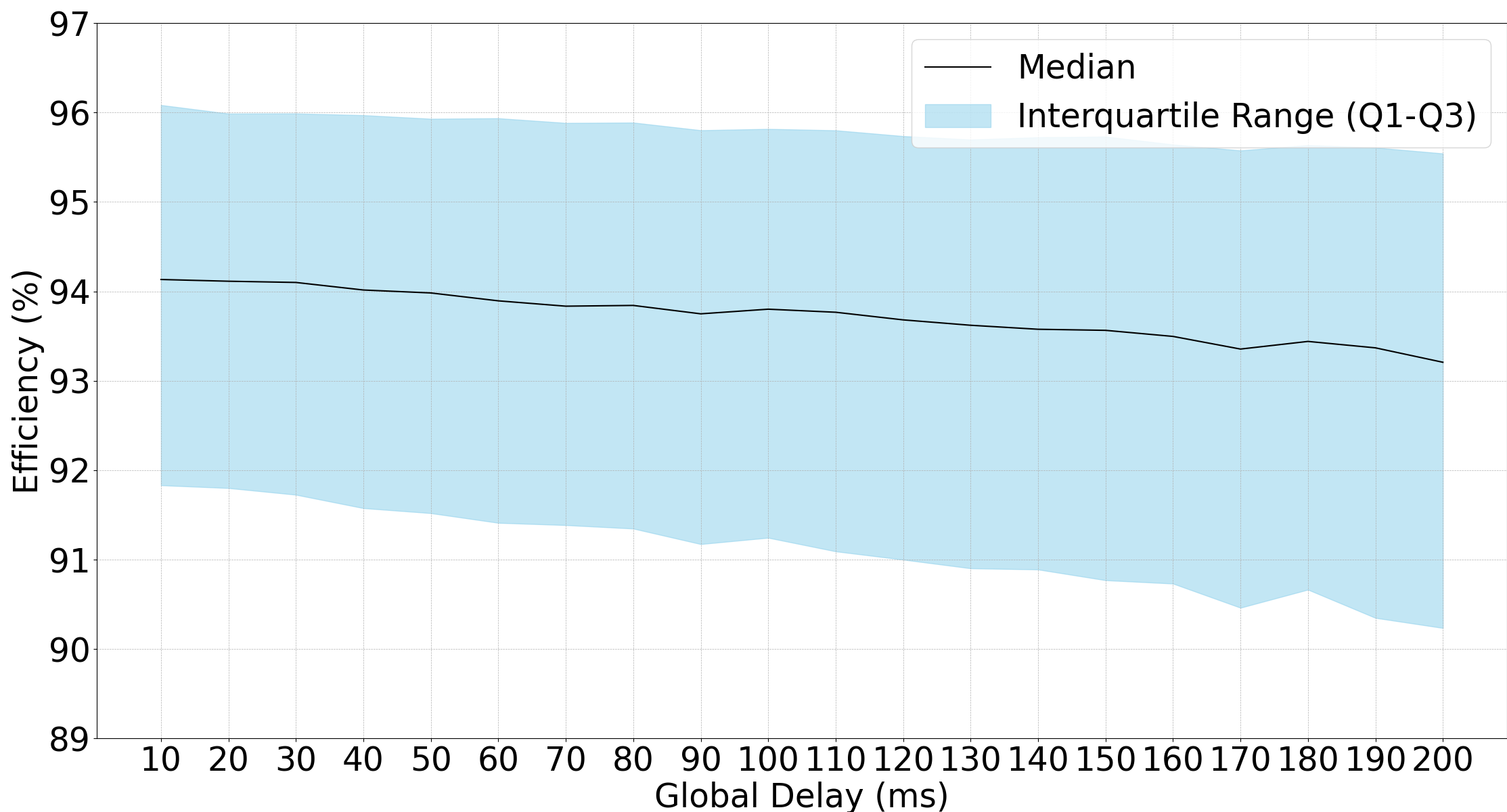}}
\caption{Auction efficiency distribution under varying global delays.}
\label{fig:eff}
\vspace{-0.2cm}\end{figure}

To evaluate the auction outcome, we introduce a measure called \emph{auction efficiency}, representing the ratio between the winning bid value and the total signal. This metric quantifies how much the auction trading mechanism allows the proposer to extract from the market as revenue. Fig.~\ref{fig:eff} demonstrates that auction efficiency gradually decreases at an average rate of 0.07\% for every 10ms increase in global delays, likely due to late-arriving MEV opportunities being excluded. In scenarios with high delays, the auction may end before builders can update bids with new profitable transactions or bundles emerging toward the auction termination. This results in a lower proportion of the total signal being included in the block, reducing auction efficiency. This analysis shows how global delay improvements increase the auction efficiency and benefit the proposer, and explains the wide adoption of optimistic relays by proposers.

\section{EOF versus Latency}
\label{sec:eof}
We conclude our analysis by comparing the effects between EOF access, i.e., probability $\pi_i$, and latency on bidding performance. To better illustrate the impact of EOF access and eliminate strategy-related effects, we consider a profile of 10 naive players with varied EOF access probabilities ranging from 0.8 to 0.98 in 0.02 increments, as exclusive transactions represent between $25\%$ and $35\%$ of the total transactions \cite{bbp}. To control for latency factors, we set both the global delay and all players' individual delay to 10ms. When comparing EOF and latency, we create an additional profile that contains 10 naive players with varied individual delays ranging from 10ms to 100ms, but with the same EOF access probability of 0.8. We run 10,000 auctions for both profiles to ensure robust results. 

\begin{figure}[t]
\centering
{\includegraphics[width=0.45\linewidth]{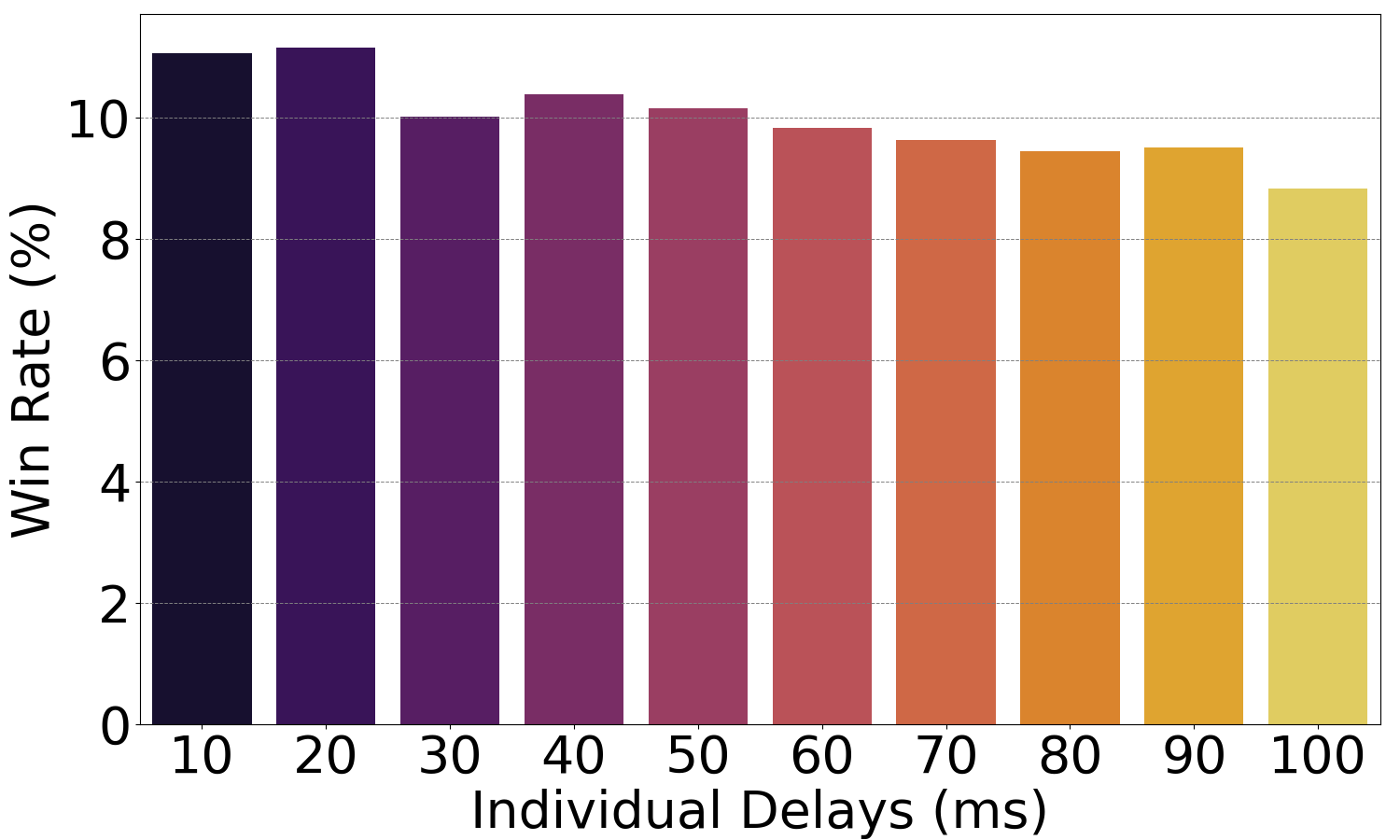}}\hspace{2pt}
{\includegraphics[width=0.45\linewidth]{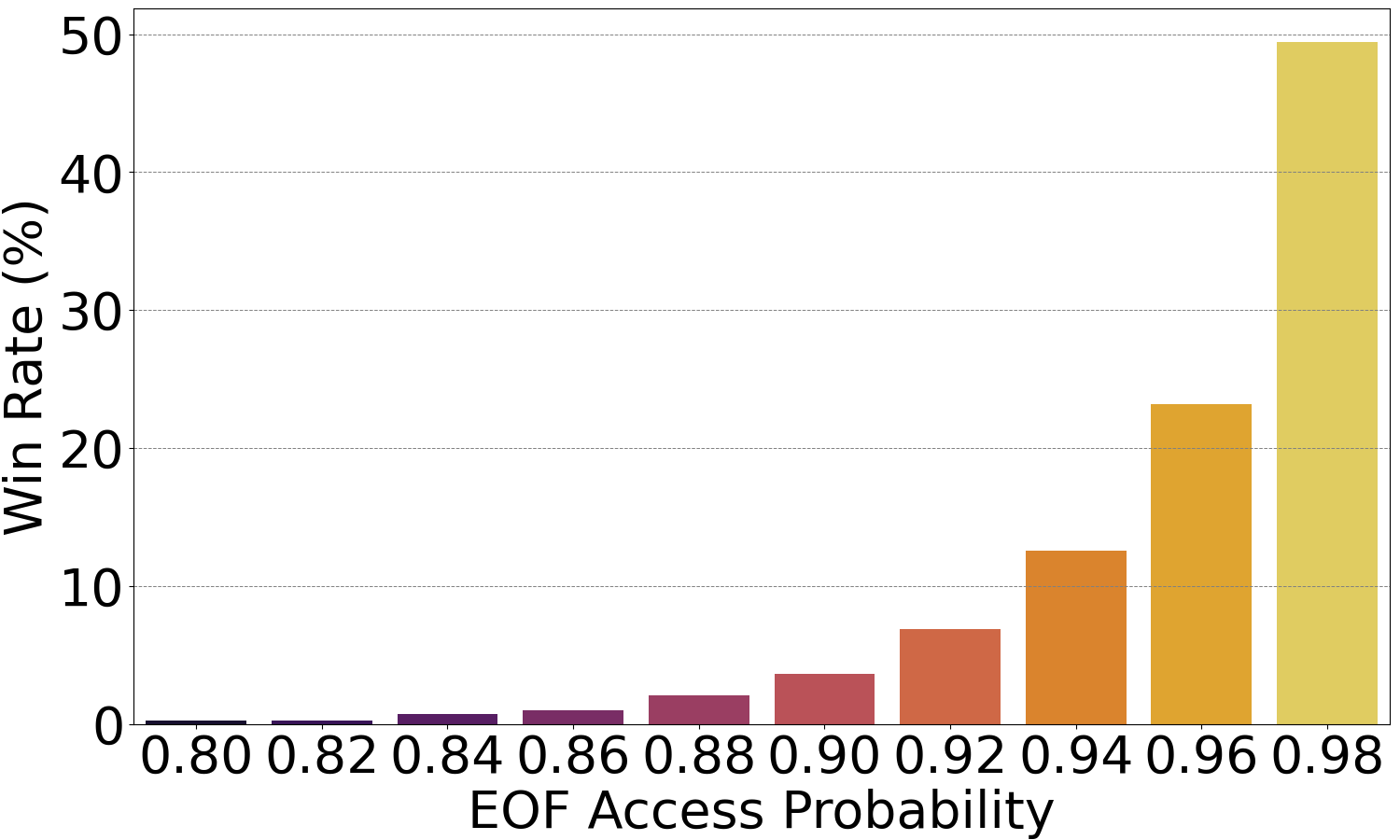}}
{\includegraphics[width=0.45\linewidth]{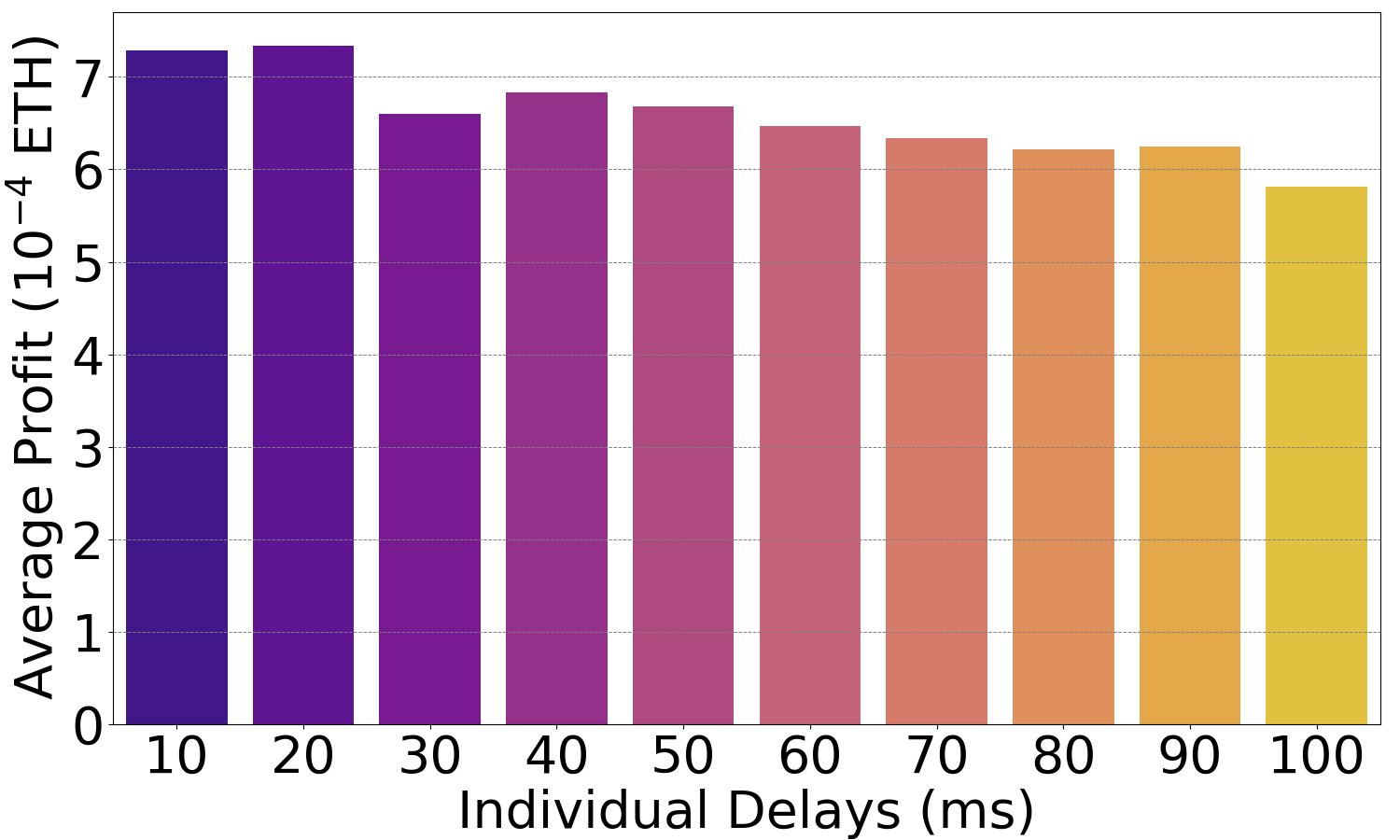}}\hspace{2pt}
{\includegraphics[width=0.45\linewidth]{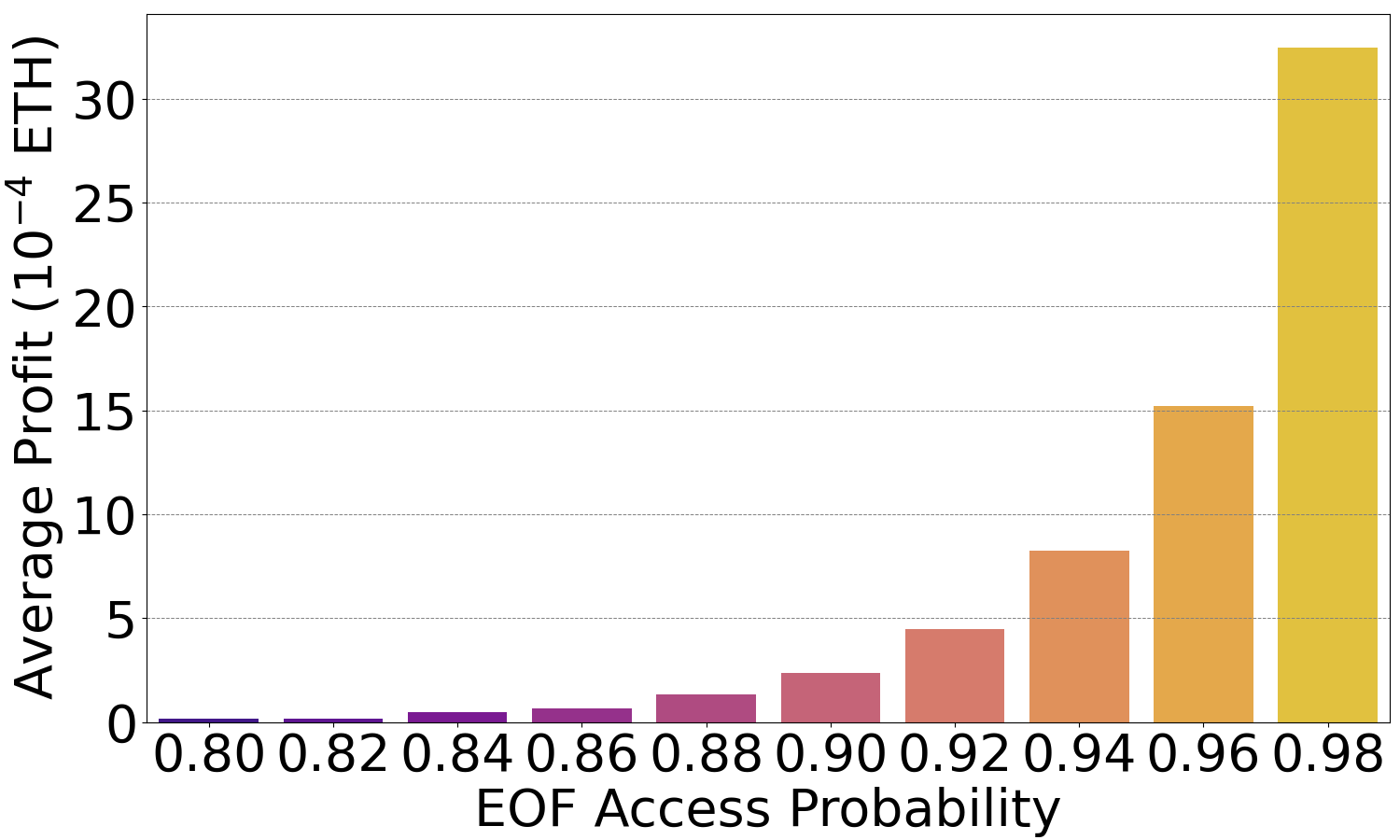}}
\caption{Impact of latency (left) vs EOF (right) on bidding performance.}
\label{fig:eof}
\vspace{-0.2cm}\end{figure}

In the left two plots of Fig.~\ref{fig:eof}, a 10ms advantage in latency contributes to an advantage of 0.26\% in win rate and $1.6 \times 10^{-5}$ ETH in average profit, except for exceptional cases where players with a higher latency outperform those with a lower latency due to possibly random access to high-value EOF. However, in the right two plots, we observe an exponential increase in players' performance as their EOF access probability increases. This provides evidence that EOF access can have a more significant impact on players' bidding performance than latency, emphasizing the importance for builders to connect with searchers to secure more EOF.

\section{Related Works}
To the best of our knowledge, MEV-Boost auctions and builders' strategic bidding behavior post Ethereum's switch to PoS remain unexplored. From the scarce literature on the topic, most closely related to our work is the study of auction games with searchers as bidders and miners (proposers) as auctioneers in the centralized MEV relay model that was initiated in \cite{qin2022quantifying}. The application of machine learning to optimize bidding strategies and predict winning bids in MEV auctions is explored in \cite{searcherauction}. A positive correlation between bid arrival time and value in the MEV-Boost block construction process is found in \cite{wahrstatter2023time}. The proposers' strategic behavior in MEV-Boost auctions is first studied in \cite{schwarz2023time} and \cite{oz2023time}, who analyze timing games in which validators strategically delay their block proposal to optimize MEV extraction. While these papers present an analysis of the impact of MEV-Boost and validator behaviors on the PoS-Ethereum block construction market, our paper is the first to analyze builders' strategic behaviors in MEV-Boost auctions.

\section{Conclusions}
In this paper, we shed light on the strategic bidding wars that currently dominate the Ethereum block construction market. Our study introduces the first game-theoretic model along a variety of bidding strategies observed in MEV-Boost auctions, involving naive, adaptive, last-minute, and bluff bidding. We study various strategic interactions and MEV-Boost auction setups and evaluate how the interplay between various critical aspects, such as access to MEV opportunities and improved connectivity to relays,  impact bidding performance. Amongst others, our findings emphasize the integral role of reducing latency to optimize builders' performance and explain why both builders and proposers favor optimistic relays, establishing a benchmark for future studies in the evolving PBS landscape.

\section*{Acknowledgment}
The authors appreciate helpful comments on the model provided by Julian Ma, Barnab\'e Monnot, Davide Crapis, and Michael Neuder, all from Ethereum Foundation. We also wish to thank Anton Wahrst{\"a}tter and his amazing project mevboost.pics for providing helpful data.

\bibliography{bib/IEEEabrv.bib, bib/IEEEexample.bib}{}
\bibliographystyle{IEEEtran}

\end{document}